\documentclass[11pt]{article}

\usepackage{graphicx}
\usepackage{calc}
\usepackage{rotating}
\usepackage[english]{babel}
\usepackage{graphicx}
\usepackage{subfig}
\usepackage{float}
\usepackage{amsmath}
\usepackage{amssymb}
\usepackage{amsthm}
\usepackage{latexsym}
\usepackage{dcolumn}
\usepackage{geometry}
\usepackage{color}
\usepackage{hyperref}
\usepackage{mciteplus}
\usepackage{slashed}
\usepackage{multirow}

\def\gtwid{\mathrel{\raise.3ex\hbox{$>$\kern-.75em\lower1ex\hbox{$\sim
$}}}}
\def\vio{\mathrel{\hbox{$E$\kern-.60em\hbox{$/
$}}}}
\newcommand{\hobs}{\ensuremath{h_{\rm obs}}}

\begin{document}

\begin{center}
{\Large \bf {Quantum interference effects in Higgs boson pair-production beyond the Standard Model} \\
\vspace*{0.8cm}
{\large Biswaranjan Das$^{a,b}$, Stefano Moretti$^c$, 
Shoaib Munir$^a$ and Poulose Poulose$^{d,e}$ } \\[0.25cm]
{\small \sl $^a$East African Institute for Fundamental Research (ICTP-EAIFR), \\ 
University of Rwanda, Kigali, Rwanda} \\[0.25cm]
{\small \sl $^b$Center for Fundamental Physics, Zewail City of Science and Technology,\\
6 October City, Giza, Egypt} \\[0.25cm]
{\small \sl $^c$School of Physics \& Astronomy,
University of Southampton, \\ Highfield, Southampton 
SO17 1BJ, UK} \\[0.25cm]
{\small \sl $^d$Department of Physics, IIT Guwahati, Assam 781039, India} \\[0.25cm]
{\small \sl $^e$Department of Physics, Concordia University, \\ 
 Montreal, QC  H4B 1R6, Canada}\bigskip \\
{\small \url{bdas@zewailcity.edu.eg}, \url{s.moretti@soton.ac.uk},
  \\ \url{smunir@eaifr.org}, \url{poulose@iitg.ac.in}}}
\end{center}
\vspace*{0.4cm}


\begin{abstract}
\noindent
New physics frameworks like the Next-to-Minimal Supersymmetric Standard Model and the Next-to-2-Higgs-doublet Model contain three neutral CP-even Higgs bosons. It is possible for the heavier two of these states to have masses identical to each other, which can result in a sizeable quantum interference between their propagators in processes they mediate. For both these models, we study the impact of such interference on the pair-production of the lightest of the three scalars, which we identify with the observed 125\,GeV Higgs boson, in the gluon-fusion channel at the Large Hadron Collider (LHC). We find that the inclusion of these effects can substantially alter the cross section, compared to its value when they are ignored, for this process. Our results illustrate the importance of taking possible quantum interference effects into account not only when investigating the phenomenology of extended Higgs sectors at the future Run(s) of the LHC, but also when imposing its current exclusion bounds on the parameter spaces of these models.
\end{abstract}


\newpage
\section{Introduction}
\label{sec:intro}

Pair-production of the Higgs boson state, \hobs, discovered in 2012 \cite{Aad:2012tfa,Chatrchyan:2012xdj} is a key process for measuring the Higgs self-coupling at the Run 3 of the Large Hadron Collider (LHC), as well as at its now 
approved high-luminosity upgrade (HL-LHC). This process represents a
direct probe of the  mechanism of electroweak symmetry breaking 
(EWSB), since the Higgs self-coupling enters the Higgs potential directly. Accessing 
it experimentally will thus be of extreme importance in order to 
understand whether mass generation in nature occurs within the 
Standard Model (SM) or in some scenario incorporating new physics.

In a beyond-the-SM (BSM) framework containing an extended Higgs sector, the phenomenology of the pair-production process of the \hobs\ candidate, i.e., the Higgs boson with mass lying near 125\,GeV, can deviate significantly from that in the SM due to two main reasons. First, the \hobs\ self-coupling gets modified from its predicted value in the SM owing to the mixing between various interaction eigenstates. Secondly, the additional Higgs states also enter the resonant channel, so that the other Higgs trilinear couplings appearing in the Lagrangian of the model also come into play. While there exists 
plenty of literature on \hobs\ pair-production in BSM scenarios 
at the various energy and luminosity stages of the LHC, most often 
this is limited to frameworks wherein only one CP-even companion to 
the SM-like Higgs state exists, like the 2-Higgs Doublet Model 
(2HDM)~\cite{Branco:2011iw} or the Minimal Supersymmetric Standard 
Model (MSSM)~\cite{Gunion:1989we,Gunion:1992hs}.

In the MSSM, the requirement for the lighter of its two scalar states, $H_1$, to be a good \hobs\ candidate pushes the heavier scalar, $H_2$, as well as the solo pseudoscalar, $A$, up into the so-called decoupling regime~\cite{Djouadi:2005gj}, where they have identical masses.\footnote{Alternatively, the $H_1$ can have SM-like properties in the `alignment without decoupling' scenario~\cite{Gunion:2002zf,Carena:2013ooa,Haber:2017erd} also.} If the MSSM Higgs sector is CP-violating, all the interaction eigenstates can mix together to yield three CP-indefinite physical states, with the two nearly mass-degenerate heavy states now labelled $H_2$ and $H_3$. When the mass-splitting between these two is comparable to the sum of their widths, a description of the intervening propagators which takes into account the imaginary parts of the one-loop self-energies, alongside the customary real parts, becomes necessary~\cite{Ellis:2004fs,Fuchs:2014ola}. This is because the imaginary off-diagonal entries of the Higgs propagator matrix can induce quantum interference between these states, so that the one produced in, for example, gluon-fusion can potentially oscillates into the other one before decaying into a given SM final state. This can significantly alter not only the total production cross section but also the shape of the differential cross section distribution for that final state~\cite{Fuchs:2016swt,Fuchs:2017wkq}. 

In the 2-Higgs-doublet model (2HDM), obtained by simply adding an additional Higgs doublet to the SM, which results again in two scalar and one pseudoscalar Higgs states, a mass-degeneracy between $H_2$ and $A$ is not a precondition for the $H_1$ to have properties identical to the \hobs. It is nonetheless a possibility not ruled out by any experimental results, and the aforementioned interference effects can become significant in this model also if it has a CP-violating Higgs sector with $m_{H_2}\approx m_{H_3}$. In a BSM scenario containing three or more CP-even Higgs bosons, the quantum interference effects can appear in processes involving Higgs propagators without the need to invoke CP-violation. A minimal realisation of such a scenario would be the extension of the two models mentioned above by a singlet Higgs field, resulting in an extra scalar state in their Higgs sectors. 

In the context of Supersymmetry, adding a complex singlet Higgs field to the MSSM can address some of its theoretical and phenomenological shortcomings, resulting in the so-called Next-to-MSSM (NMSSM)~\cite{Fayet:1974pd,Ellis:1988er,Durand:1988rg,Drees:1988fc}. In this model, some particular configurations of the free parameters can yield a SM-like $H_1$ along with $H_2$ and $H_3$ that are nearly mass-degenerate. We have previously investigated the aforementioned interference effects in the NMSSM, in the scenario where $H_1$ and $H_2$ are mass-degenerate~\cite{Das:2017tob}, as well as in the alternative scenario with $m_{H_2}\approx m_{H_3}$~\cite{Das:2018haz}. The first study pertained to the production process for the $\gamma\gamma$ final state and the second to that of $\tau^+\tau^-$ at the LHC. Both these studies found the results from the calculation embedding the full Higgs propagator matrix to be notably different from the ones using the standard approximation where only one term containing a Breit-Wigner (BW) propagator corresponding to each of the Higgs bosons appears in the amplitude expression. It was also shown in those papers that the expected mass resolutions of the respective final states at the LHC may, however, not allow it to disentangle the two Higgs states from each other.   

In this article, we investigate the implications of the quantum interference on the gluon-initiated pair-production of the SM-like $H_1$ state of the NMSSM at the LHC with $\sqrt{s}=14$\,TeV, and also of its non-Supersymmetric counterpart, the Next-to-2HDM (N2HDM). The latter model is obtained by introducing a real singlet Higgs field into the 2HDM, and while it is phenomenologically similar to the NMSSM, a crucial advantage the N2HDM has is that the physical Higgs boson masses can themselves be the input parameters. This grants us the freedom of setting the $H_2$ and $H_3$ masses exactly equal and assessing the impact of this maximal mass-degeneracy on the said process. This model additionally allows us to analyse how the various Higgs couplings govern the relative sizes of the interference effects, so that the general inferences can be extended to other multi-Higgs BSM scenarios.

The article is organised as follows. In the next section we briefly revisit the Higgs pair-production process at the LHC. In section \ref{sec:models} we discuss some details of the NMSSM and the N2HDM, as well as of our numerical computational tool. In section \ref{sec:numeric} we present our analysis and discuss its results. In section~\ref{sec:concl} we conclude our findings.


\section{Higgs pair-production at the LHC}
\label{sec:pair-prod}

\begin{figure}[t!]
\centering\begin{tabular}{cc}
\includegraphics*[scale=1.1]{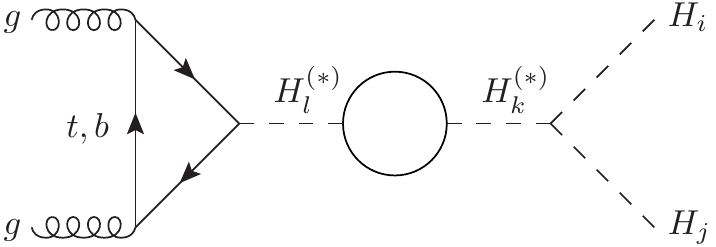} &
\hspace*{1.0cm}\includegraphics*[scale=1.1]{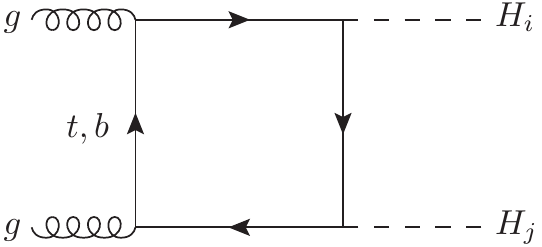} \\
\end{tabular}
\caption{\label{fig:lambda-width} The Feynman diagrams contributing to Higgs boson pair-production in a model with an extended Higgs sector, but no additional particle content besides the SM.}
\label{fig:FDs}
\end{figure}

The cross section for the (inclusive) process $pp \to H_iH_j$, where $i,j=1,...,N$ for a model with $N$ CP-even Higgs bosons but without any additional particle content beyond the SM, can be written at the leading order (LO) as
\begin{equation}\label{eq:diffXShats}
\sigma_\text{LO}(pp\to H_iH_j) =
\int_{0}^{1} d\tau \int_{\tau}^{1} \frac{dx_1}{x_1}
 \frac{g(x_1) g(\tau/x_1)}{1024 \pi \hat s^3} 
{\cal A}_{gg \to H_iH_j}^2\,,
\end{equation}
where $g(x_1)$ and $g(x_2)$ are the parton distribution functions (PDFs) of the two incoming gluons having squared centre-of-mass (CM) energy ${\hat s} = x_1 x_2 s$, given in terms of the total CM energy, $s$, of the $pp$ system, and by defining $\tau \equiv \frac{\hat s}{s} = x_1 x_2$. The amplitude-squared in Eq.\,(\ref{eq:diffXShats}) can be written, following the notation of Ref.\,\cite{Plehn:1996wb}, as 
\begin{equation}
\label{eq:amplitude}
{\cal A}^2_{gg \to H_iH_j}=\Big|C_\vartriangle F_\vartriangle + C_\Box F_\Box \Big|^2 + \Big|C_\Box G_\Box \Big|^2,
\end{equation}
where $\vartriangle$ denotes the contribution from the Higgs-mediated triangle loop diagram, Fig.\,\ref{fig:FDs} (left), and $\Box$ refers to the quark-box diagram, Fig.\,\ref{fig:FDs} (right). 

The coefficient corresponding to the box contributions in Eq.\,(\ref{eq:amplitude}) is written in terms of the Yukawa couplings as
\begin{equation}
C_\Box = \displaystyle\sum_{q} g_{H_i\bar{q}q}g_{H_j\bar{q}q}\,.
\end{equation}
The form factor $F_\Box$ corresponds to the case when the gluons have a combined total spin of $S_z = 0$ along the proton beam, while $G_\Box$ refers to the case with $S_z = 2$. The full expressions for $F_\Box$ and $G_\Box$ within the SM can be found in the appendix of Ref.\,\cite{Plehn:1996wb}. 

The Higgs-mediated triangle loop diagram contributes only to the $S_z = 0$ case. The corresponding form factor, for state $H_l$ attached to the triangle, is written as
\begin{equation}\label{eq:ftri}
F^l_\vartriangle = \frac{\alpha_s \hat s}{4 \pi v} \Bigl\{ S_l^g + i \lambda P_l^g\Bigr\}\,, 
\end{equation}
where the scalar and pseudoscalar components, $S_l^g$ and $P_l^g$, 
respectively, can be found in, e.g., Refs.\,\cite{Lee:2003nta,Baglio:2013iia}. In case of a single Higgs boson, as in the SM, the triangle coefficient in Eq.\,(\ref{eq:amplitude}) is given as
\begin{equation}\label{eq:ctribw}
C_\vartriangle =  \frac{M_Z^2}{\hat s - M_h^2}~\lambda_{hhh},
\end{equation}
where $\lambda_{hhh}$ is the Higgs trilinear self-coupling. In multi-Higgs models like the ones we intend to explore here, the above coefficient is generalised to 
\begin{equation}\label{eq:ctri}
C^l_\vartriangle \equiv \displaystyle\sum_{k=1}^N {\cal D}_{kl}(\hat s)\lambda_{H_iH_jH_k}\,.
\end{equation}
Here, $\lambda_{H_iH_jH_k}$ are the Higgs trilinear couplings and ${\cal D}_{kl}(\hat s)$, with $k,l=1,...,N$, are the entries of the Higgs propagator matrix. This modified $C^l_\vartriangle$ allows the possibility of interference between two different Higgs intermediate states, induced by  higher order quantum effects, as illustrated by Fig.\,\ref{fig:FDs} (left). 

The main focus of this study is to investigate the above-mentioned quantum effects in the specific scenario with $N=3$, which permits the resonant pair-production of the lightest Higgs state via the two,  mutually interfering, heavier states. In this scenario, the (symmetric) propagator matrix is written as
\begin{equation}
\label{eq:propmat}
{\cal D}({\hat s}) = \hat s 
\left(\begin{array}{@{}ccc@{}} 
{\hat s}-m_{H_1}^2 +i{{\mathfrak{I}}{\rm m}\hat\Pi}_{11}(\hat s)
& i{{\mathfrak{I}}{\rm m}\hat\Pi}_{12}(\hat s) 
& i{{\mathfrak{I}}{\rm m}\hat\Pi}_{13}(\hat s)\\
i{{\mathfrak{I}}{\rm m}\hat\Pi}_{21}(\hat s) 
& {\hat s}-m_{H_2}^2 +i{{\mathfrak{I}}{\rm m}\hat\Pi}_{22}(\hat s)
& i{{\mathfrak{I}}{\rm m}\hat\Pi}_{23}(\hat s) \\
i{{\mathfrak{I}}{\rm m}\hat\Pi}_{31}(\hat s) 
& i{{\mathfrak{I}}{\rm m}\hat\Pi}_{32}(\hat s) 
& {\hat s}-m_{H_3}^2 +i{{\mathfrak{I}}{\rm m}\hat\Pi}_{33}(\hat s)\\
\end{array}\right)^{-1}\,, 
\end{equation}
where the ${{\mathfrak{I}}{\rm m}\hat\Pi}_{kl}(\hat s)$ are the absorptive parts of the Higgs self-energies, and $m_{H_k}$ is the renormalised mass of the $k$-th Higgs boson. The explicit expressions for ${{\mathfrak{I}}{\rm m}\hat\Pi}_{kl}(\hat s)$ can be found in the Appendix of Ref.\,\cite{Das:2017tob}. In general, however, the off-diagonal absorptive terms in the propagator are assumed to be negligible, in which case the ${\cal D}({\hat s})$ becomes a diagonal matrix and $C_\vartriangle$ can, to a good approximation, be reduced to a sum over three terms containing BW propagators corresponding to each $H_l$, as in Eq.\,(\ref{eq:ctribw}). 

\section{Models and computational tools}
\label{sec:models}

Two new physics models that are consistent with the $N=3$ scenario are the NMSSM and  N2HDM. In both these models, we identified the lightest of the three scalars, $H_1$, with $h_{\rm obs}$, and analysed the impact of the interference between the heavier $H_2$ and $H_3$ on its pair-production at the $\sqrt{s}=14$\,TeV LHC. It has previously been emphasised in literature ~\cite{Fuchs:2014ola,Das:2017tob,Carena:2000yi,Cacciapaglia:2009ic} that these effects are more pronounced for large (combined) total widths of the intermediate Higgs states compared to the mass splitting between these. Using this as a guideline, we  numerically scanned the parameter spaces of the two models to find their potentially relevant configurations. Below we discuss some details of the two models as well as of our calculation of the $H_1H_1$ production cross section.


\subsection{NMSSM}
\label{sec:nmssm}

As a follow-up of our previous analyses, the first model that we investigate is the $Z_3$-symmetric NMSSM. The Higgs potential in this model is written in terms of the two $SU(2)_L$ doublets $H_u$ and $H_d$, with $Y = \pm 1$, and the singlet $S$ as
\begin{eqnarray}\label{eq:potential}
V_{\rm NMSSM} &=& {| \lambda \left(H_u^+ H_d^- - H^0_u H^0_d \right) + \kappa S^2 |}^2
+ m_S^2 {| S |}^2 + \left(m_{H_u}^2 + {| \lambda S |}^2 \right)
  \left({| H^0_u |}^2 + {| H_u^+ |}^2 \right) \nonumber \\ 
&+& \left(m_{H^0_d}^2 + {|\lambda S |}^2 \right)
  \left({| H^0_d |}^2 + {| H_d^- |}^2 \right)
+ \frac{g_1^2+g_2^2}{8} \left({| H^0_u |}^2 + {| H_u^+ |}^2
- {| H^0_d |}^2 - {| H_d^- |}^2 \right)^2 \nonumber \\
&+& \frac{g_2^2}{2} {| H_u^+ H_d^{0*} + H^0_u H_d^{-*} |}^2
+ \left[ \lambda A_{\lambda} \left(H_u^+ H_d^- - H^0_u H^0_d \right) S
+ \frac{1}{3} \kappa A_{\kappa} S^3 + {\rm h.c.} \right].
\end{eqnarray}
Here $\lambda$ and $\kappa$ are dimensionless Higgs trilinear 
couplings and $A_{\lambda}$ and $A_{\kappa}$ are
their respective soft SUSY-breaking counterparts, 
$m_{H_d}$, $m_{H_u}$ and $m_S$ are the soft Higgs masses, 
while $g_1$ and $g_2$ are the $U(1)_Y$ and $SU(2)_L$ gauge 
coupling constants, respectively.

The neutral components of the fields $H_d$, $H_u$ and $S$ are developed around their respective vacuum expectation values (VEVs) $v_d$, $v_u$ and $v_S$, when EW symmetry is broken, as
\begin{equation}\label{eq:fields}
H^0_d = 
\left( \begin{array}{c} v_d+H_{dR}+iH_{dI} \\ H_d^- \end{array} \right),~~~
H^0_u =
\left( \begin{array}{c} H_u^+ \\ v_u+H_{uR}+iH_{uI} \end{array} \right),~~~
S =  v_S+S_R+iS_I\,.
\end{equation}
By taking the second derivative of $V_{\rm NMSSM}$, one then obtains the tree-level $3 \times 3$ neutral CP-even Higgs mass-squared matrix, $M_H^2$, in the $(H_{dR},H_{uR},S_R)^T$ basis. The orthogonal matrix ${\cal R}$ rotates these interaction eigenstates into the physical states as
\begin{equation}\label{eq:mdiagn2hdm}
\left( H_1,H_2,H_3 \right)^T={\cal R} \left(H_{dR},H_{uR},S_R \right)^T\,.
\end{equation}
The matrix $M_H^2$ thus gets diagonalised as
\begin{equation}\label{eq:mdiagn2hdm2}
{\cal R}M_H^2{\cal R}^T={\rm diag}\left(m_{H_1}^2,m_{H_2}^2,m_{H_3}^2\right),
\end{equation}
with Higgs boson masses in the ascending order, i.e., $m_{H_1}<m_{H_2}<m_{H_3}$.

Our current analysis pertains to the `phenomenological' NMSSM, wherein all the free parameters, including the above Higgs sector ones, are input at the EW scale. Since variations in non-Higgs sector parameters are expected to have little impact on our particular phenomenological scenario, we fixed the soft squark masses as $M_{Q_{1,2,3}} = M_{U_{1,2,3}} = M_{D_{1,2,3}} = 3$\,TeV, the slepton masses as $M_{L_{1,2,3}} = M_{E_{1,2,3}}= 2$\,TeV, the soft gaugino masses as $2M_1 = M_2 = \frac{1}{3}M_3 = 1$\,TeV. This resulted in $\tan\beta$ $(\equiv \frac{v_u}{v_d})$, $\mu_{\rm eff}$ $(\equiv \lambda v_s)$, $\lambda$, $\kappa$, $m_P$, $m_A$, and the unified trilinear coupling of the charged sfermions, $A_0 \equiv A_{\tilde{u},\tilde{c},\tilde{t}} = A_{\tilde{d},\tilde{s},\tilde{b}} = A_{\tilde{e},\tilde{\mu},\tilde{\tau}}$, as the complete set of inputs. The parameters $m_P$ and $m_A$ are the bare masses of the two pseudoscalars, which are a trade-off for $A_\lambda$ and $A_\kappa$ using the minimisation conditions of the Higgs potential.

\begin{table}[tbp]
\begin{center}
\begin{tabular}{|c|c|c|}
\hline
 Parameter & Scanned range & Range giving $m_{H_{2,3}}\leq 500$\,GeV  \\
\hline
$A_0$\,(GeV) & $-5000$ -- 0 & $-5000$ -- $-3500$ \\
$\mu_{\rm eff}$\,(GeV) & 100 -- 1000 & 100 -- 250 \\
$\tan\beta$ & 1 -- 40 & 5 -- 10 \\
$\lambda$  & 0.001 -- 0.7 & 0.001 -- 0.3  \\
$\kappa$  & 0.001 -- 0.7 & 0.001 -- 0.5 \\
$m_P$\,(GeV) & $100$ -- 1000 & 100 -- 500 \\
$m_A$\,(GeV) & $100$ -- 1000 & 400 -- 500 \\
\hline
\end{tabular}
\caption{Ranges of the NMSSM input parameters scanned for obtaining $H_2$ and $H_3$ with large mass-degeneracy. The third column shows the parameter space yielding $m_{H_{2,3}}\leq 500$\,GeV.}
\label{tab:params}
\end{center}
\end{table}

We used the public code {\tt NMSSMTools-v5.5.2}~\cite{NMSSMTools,Ellwanger:2004xm,Ellwanger:2005dv} for numerically generating the Higgs boson mass spectra and branching ratios (BRs) corresponding to each set of values of the 7 model input parameters, randomly selected from the ranges shown in the second column of Table\,1. Each parameter space point was required to satisfy all the theoretical and experimental constraints defined in {\tt NMSSMTools}, which include limits from the Higgs searches at the Large Electron-Positron (LEP) collider, the TeVatron and the LHC, from the direct and indirect searches for neutralino dark matter (DM) and estimates of its relic abundance and from $B$-physics measurements. In our scenario, since the $H_1$ plays the role of the $h_{\rm obs}$, {\tt NMSSMTools} intrinsically imposes 2$\sigma$ bounds on its couplings from the most relevant recent LHC results, while also requiring $m_{H_1}$ to lie within the $122 - 128$\,GeV range, allowing a $\pm 3$\,GeV theoretical uncertainty on the measured mass of $\sim 125$\,GeV. Output points satisfying all these constraints were further run through {\tt HiggsBounds-v5.7.0}~\cite{Bechtle:2008jh,Bechtle:2011sb,Bechtle:2013wla,Bechtle:2015pma,Bechtle:2020pkv} to test the Higgs sector observables of the model against the latest exclusion bounds from the LHC that might not (yet) have been included in {\tt NMSSTools} itself.  

\begin{figure}[t!]
\centering\includegraphics*[width=12cm]{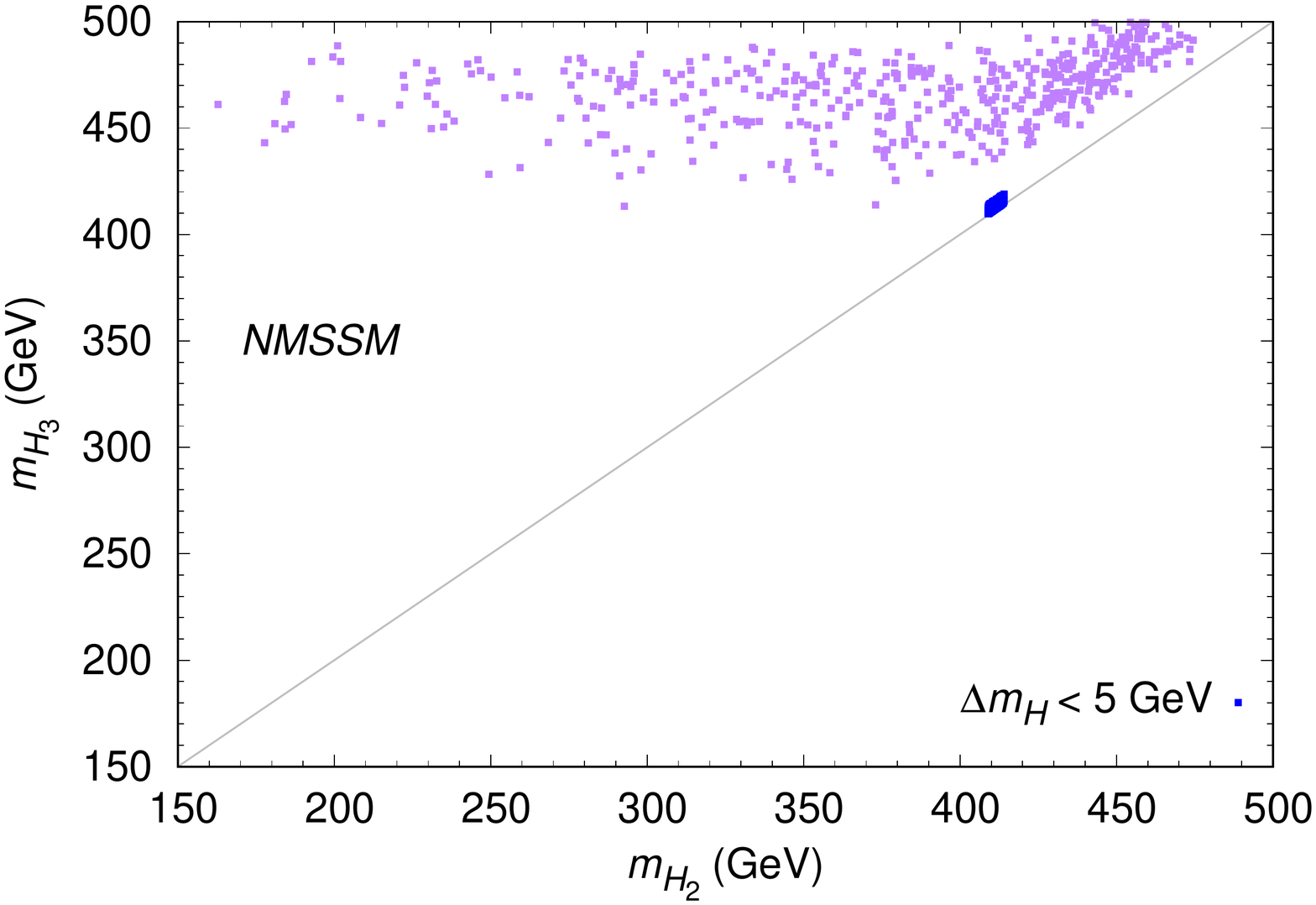} 
\vspace*{-0.8cm}\caption{\label{fig:NM-mh23} The masses of $H_2$ and $H_3$ for the points obtained in the scans of the extended parameter space of the NMSSM (violet) and of its narrow region yielding $\Delta m_H<5$\,GeV (blue).}
\end{figure}

In Fig.\,\ref{fig:NM-mh23} we show the resulting successful points with $m_{H_{2,3}}\leq 500$\,GeV, which are obtained for the input parameter ranges given in the third column of Table\,1. One notices that the limits from the direct searches at the LHC rule out a mass below $\sim 405$ GeV for the (predominantly doublet-like) $H_3$, over the entire parameter space explored. Our initial scan with wide input ranges of the parameters yielded only one point (out of nearly two thousand violet points in the figure) with $\Delta m_H \equiv m_{H_3}-m_{H_2}$ less than 5\,GeV, lying just above the LHC exclusion bound for $m_{H_3}$. In order to find solutions with larger $H_2\text{-}H_3$ mass-degeneracy, we therefore performed another scan of the narrowed-down parameter space region around the said point. Indeed, several points with $\Delta m_H <1$\,GeV were obtained with this secondary scan, which are plotted in blue colour in the figure. The coordinates of the point with the smallest $\Delta m_H$ are 
\begin{gather}\label{eq:inputnmssm}
\tan\beta=6,~~A_0=-5000\,{\rm GeV},~~\lambda=0.005,~~\kappa=0.0071,  \nonumber \\
\mu_{\rm eff}=148.24\,{\rm GeV},~~m_P=147.59\,{\rm GeV},~~m_A=431.25\,{\rm GeV},
\end{gather}
which result in the following Higgs mass spectrum:
\begin{gather}
m_{H_1}=122.23\,{\rm GeV},~~m_{H_2}=409.33\,{\rm GeV},~~m_{H_3}=410.13\,{\rm GeV},\nonumber \\ 
m_{A_1}=147.59\,{\rm GeV},~~m_{A_2}=408.23\,{\rm GeV},~~m_{H^\pm}=416.13\,{\rm GeV}.
\label{eq:masses-nmssm} 
\end{gather}

The total widths of the three scalars yielded by the above parameter space point are $\Gamma_{H_1}=4.76$\,MeV, $\Gamma_{H_2}=535.4$\,MeV and $\Gamma_{H_3}=24.78$\,MeV. $H_2$ in this point is doublet-like, while $H_3$ is singlet-like. For this reason, the latter has much weaker couplings to the SM, and hence much smaller partial decay widths, than the former. We also point out that for many points obtained in the initial wider scan, $m_{H_1}$ easily reaches up to 125\,GeV. Its lying close to the enforced lower limit of 122\,GeV for the point in Eq.\,(\ref{eq:masses-nmssm}) is simply a consequence of the very narrow parameter space scanned to obtain maximally degenerate $m_{H_2}$ and $m_{H_3}$, especially with the soft squark and gaugino masses fixed. For this point, the $A_0$ parameter, larger magnitudes of which were generally preferred by the points in order to push $m_{H_1}$ above 122\,GeV, sits right at the upper end of its scanned range. 

\subsection{N2HDM}
\label{sec:n2hdm}

Since the scans for the NMSSM did not generate any points with a $H_3$ lighter than $\sim 405$\,GeV, we extended our analysis to the N2HDM also. In this model the physical masses of the three scalar Higgs bosons, $m_{H_i}$, are input parameters, as opposed to the NMSSM, wherein they are derived quantities, which allows greater freedom in the selection of the other free parameters relevant to the process under investigation. The N2HDM is obtained by adding a real singlet scalar field, $S$, to the (CP-conserving) 2HDM, and its Higgs potential reads
\begin{eqnarray}\label{eq:potentialn2hdm}
V_{\rm N2HDM} &=& m_{H_u}^2|H_u|^2+m_{H_d}^2|H_d|^2-m_{12}^2\left(H_u^\dagger H_d+{\rm h.c.}\right)+\frac{\lambda_1}{2}\left(H_u^\dagger H_u\right)^2+\frac{\lambda_2}{2}\left(H_d^\dagger H_d\right)^2 \nonumber \\
&+&\lambda_3\left(H_u^\dagger H_u\right)\left(H_d^\dagger H_d\right)+\lambda_4\left(H_u^\dagger H_d\right)\left(H_d^\dagger H_u\right)+\frac{\lambda_5}{2}\left\{\left(H_u^\dagger H_d\right)^2+{\rm h.c.}\right\} \nonumber \\
&+&\frac{m_S^2}{2}S^2+\frac{\lambda_6}{8}S^4+\frac{\lambda_7}{2}\left(H_u^\dagger H_u\right)S^2+\frac{\lambda_S}{2}\left(H_d^\dagger H_d\right)S^2,
\end{eqnarray}
where $H_u$ and $H_d$ are doublet fields similar to the NMSSM ones. This potential has a generic form and observes two symmetries: i) a $Z_2$-symmetry, $H_u \to H_u,\, H_d \to - H_d,\,S \to S$, which is softly broken by the term containing $m_{12}^2$, and ii) a spontaneously broken $Z_2^\prime$-symmetry, $H_u \to H_u,\, H_d \to H_d,\, S \to -S$.  

The charge assignments of the fermions under the $Z_2$ symmetry define the four \textit{types} of the underlying (N)2HDM. Our adopted notation for the doublet Higgs fields is intended to indicate the Type-II N2HDM specifically, wherein the fermions have $Z_2$ charges such that the doublet $H_u$ couples only to the up-type quarks and $H_d$ to the down-type quarks and charged leptons. Upon EWSB, the two doublet fields are expanded around their respective VEVs according to Eq.\,(\ref{eq:fields}), while the real singlet is expanded in this model as $S =  v_S+S_R$. After minimisation of the potential and rotation of the scalar mass matrix, as in the NMSSM, the masses of the three physical CP-even Higgs states are obtained, with $m_{H_1}<m_{H_2}<m_{H_3}$. Besides these, the Higgs sector of the model also contains a CP-odd Higgs boson $A$. The Type-II N2HDM is thus essentially the non-Supersymmetic counterpart of the NMSSM, with fewer symmetries impinging on the properties of the CP-even Higgs sector, which makes it a more suitable fit for our comparative investigation than the other N2HDM types. For details of the Higgs sector of the N2HDM, we refer the reader to Refs.~\cite{vonBuddenbrock:2016rmr,Muhlleitner:2016mzt}. 

There are twelve free parameters in the potential in Eq.\,(\ref{eq:potentialn2hdm}): $\lambda_{1,\cdots,~7, S},~ m_{H_u}^2,~m_{H_d}^2,~m_{S}^2,~m_{12}^2$. Relations between these parameters and the VEVs, arising from the minimisation conditions of the Higgs potential, allow us to trade $m_{H_u}^2,~m_{H_d}^2$, and $m_{S}^2$ for $\tan\beta,~v \left(\equiv \sqrt{v_u^2+v_d^2}\right)$ and $~v_S$. Moreover, the eight quartic couplings can be traded for the physical masses, $m_{H_{1,2,3}},~m_{H^\pm},~m_A$, and the three independent parameters of the mixing matrix $\mathcal{R}$ in Eq.\,(\ref{eq:mdiagn2hdm}). These parameters, taken to be $\mathcal{R}_{11},~\mathcal{R}_{12}$ and $\mathcal{R}_{23}$, can then further be replaced by the top-Yukawa and gauge couplings of the $H_1$, defined in units of the corresponding couplings of the Higgs boson in the SM as
\begin{equation}
g_{H_1tt}=\frac{{\cal R}_{12}}{\sin\beta},~~~~~~g_{H_1VV}=\cos\beta~{\cal R}_{11}+\sin\beta~{\cal R}_{12}.
\end{equation}
Thus, for the purpose of this study, the following independent real parameters representing the N2HDM were randomly scanned in the given ranges using the public tool {\tt ScannerS-2} \cite{Coimbra:2013qq,Muhlleitner:2020wwk}:
\begin{gather}
m_A:500 \text{-} 1000\,{\rm GeV},~~m_{H^\pm}:500\text{-}1000\,{\rm GeV},~~m_{12}^2:10^4\text{-}10^5\,{\rm GeV}^2,~~\tan\beta:1\text{-}20, \nonumber \\
g_{H_1VV}^2,\,g_{H_1t\bar t}^2:0.64\text{-}1.44,~~{\rm sign}({\cal R}_{13}):\pm,~~{\cal R}_{23}:-1\text{-}1,~~v_S:1500\text{-}2500\,{\rm GeV},
\label{eq:inputn2hdm2}
\end{gather}
where sign$({\cal R}_{13})$ takes into account the sign ambiguity in the neutral scalar mixing. In this model, the Higgs trilinear couplings $g_{H_2H_1H_1}$ and $g_{H_3H_1H_1}$, which are of particular relevance for the process of our interest here, are given as
\begin{eqnarray}
g_{H_jH_iH_i}&=&\frac{3}{v}\left[-\frac{1}{2}\tilde{\mu}^2\left( \frac{{\cal R}_{i2}}{\sin\beta}-\frac{{\cal R}_{i1}}{\cos\beta}\right)\left(6{\cal R}_{i2}{\cal R}_{j2}+6{\cal R}_{i3}{\cal R}_{j3}\sin^2\beta +\sum_k \epsilon_{ijk}{\cal R}_{k3}\sin 2\beta\right)\right. \nonumber \\
&+& \left. \frac{2m_{H_i}^2+m_{H_j}^2}{v_S}\left({\cal R}_{i3}^2{\cal R}_{j3}v+{\cal R}_{i2}^2{\cal R}_{j2}\frac{v_S}{\sin\beta}+{\cal R}_{i1}^2{\cal R}_{j1}\frac{v_S}{\cos\beta}\right)\right] ,
\end{eqnarray}
where $\tilde{\mu}^2\equiv\frac{m_{12}^2}{\sin\beta\cos\beta}$ and $\epsilon_{ijk}$ is the totally antisymmetric tensor, with $\epsilon_{123}=1$.

While the above ranges were mostly guided by existing literature on the model (see, e.g., Refs.~\cite{Muhlleitner:2016mzt,Krause:2017mal,Ferreira:2019iqb}), the ones of $g^2_{H_1VV}$ and $g^2_{H_1t\bar t}$ were based loosely on the current 2$\sigma$ error-bar on the measurements of the corresponding couplings for the $h_{\rm obs}$ at the LHC~\cite{Sirunyan:2018koj}. Several scans were performed for this model, in all of which we fixed $v=246$\,GeV and $m_{H_1} =125$\,GeV. The values of $m_{H_{2,3}}$, in contrast, were set to certain different values of interest in different scans (as will be explained in the next section). The purpose of the numerical scanning was to find configurations of the parameters in Eq.\,(\ref{eq:inputn2hdm2}) that satisfied theoretical conditions such as unitarity and vacuum stability, and were at the same time consistent with precision EW and $B$-physics measurements. In addition to these checks performed internally by {\tt ScannerS}, testing of the Higgs sector observables against the exclusion bounds from direct collider searches was also performed for each scanned point, by interfacing it with {\tt N2HDECAY}~\cite{Engeln:2018mbg} and {\tt HiggsBounds}. Finally, {\tt ScannerS} was also interfaced with the program {\tt HiggsSignals-2}~\cite{Bechtle:2013xfa,Bechtle:2020uwn}, which performs a $\chi^2$-fit of the \hobs\ properties for a given model point against the LHC measurements, and rules it out if $\Delta\chi^2=\chi^2_{\rm N2HDM}-\chi^2_{\rm SM}>6.18$ (assuming a $2\sigma$ Gaussian error on the best-fit value).
 

\subsection{Cross section calculation}

For the output points from the scans, we proceeded to calculate the inclusive $pp \to H_1 H_1$ cross section, using a {\tt FORTRAN} code prepared in-house. For evaluating $\sigma_{\rm LO}$ given in Eq.\,(\ref{eq:diffXShats}), the expressions corresponding to the triangle and box form factors were formulated following the public code {\tt HPAIR-v2.00}~\cite{Plehn:1996wb,Dawson:1998py,Grober:2015cwa}, which includes only the SM and the MSSM. The numerical computation of the next-to-LO (NLO) corrections to $\sigma_{\rm LO}$, which can be expressed as~\cite{Dawson:1998py}
\begin{equation}
\Delta \sigma = \Delta \sigma_\text{virt} + \Delta \sigma_{gg} + \Delta \sigma_{gq} + \Delta \sigma_{\bar{q}q}\,,
\end{equation} 
were also imported from {\tt HPAIR}, since they are generic to all models.
Besides catering to models beyond the MSSM, another significant way that our cross section calculator differs from {\tt HPAIR}, which evaluates individual BW propagators for each intermediate Higgs boson in the triangle diagram,  is in the incorporation of the
full propagator matrix of Eq.\,(\ref{eq:propmat}). This allows us to estimate the magnitude of the  effects resulting from the off-diagonal terms in the matrix, by including or neglecting these during the cross section computation for a given point by our code. 

Since the input parameters as well as the particle contents, and hence the Higgs self-energy contributions, of the NMSSM and N2HDM are mutually rather different, we prepared a separate code for each of these models. In order to check the accuracy and consistency of our base code, we compared the $pp \to H_1H_1$ cross sections calculated in the MSSM limit of the NMSSM for a few test points with the ones obtained from {\tt HPAIR}. We found the two sets of results to be in very good (within 1\%) agreement. Note here that the higher order QCD corrections for this process have now been evaluated up to the next-to-next-to-next-to-LO\,\cite{Agostini:2016vze,Degrassi:2016vss,Bonciani:2018omm,Chen:2019lzz,Chen:2019fhs,Baglio:2020ini} in the SM. We presume that these can be extended straightforwardly to the multi-Higgs models discussed here, and their overall impact would amount to a simple rescaling of our NLO calculations.


\section{Analysis results}
\label{sec:numeric}

To quantify the magnitude of the triangle-box interference arising in the $S_z=0$ channel and, additionally, the full-propagator effects within the triangle diagram, we calculated the integrated cross sections corresponding to the following cases for each successful point from the scans for the two models: \\
\noindent a) without triangle-box interference, with diagonal-only propagator matrix, \\
\noindent b) with triangle-box interference, with diagonal-only propagator matrix, \\
\noindent c) with triangle-box interference, with full propagator matrix. \\
Below, these three cross sections will be referred to as $\sigma_{\rm a}$, $\sigma_{\rm b}$, and $\sigma_{\rm c}$, respectively. We also define $R_\sigma \equiv \sigma_{\rm b}/\sigma_{\rm c}$.


\subsection{The NMSSM}
\label{subsec:num-nmssm}

\begin{figure}[t!]
\begin{tabular}{cc}
\hspace*{-1.0cm}\includegraphics*[width=9.5cm]{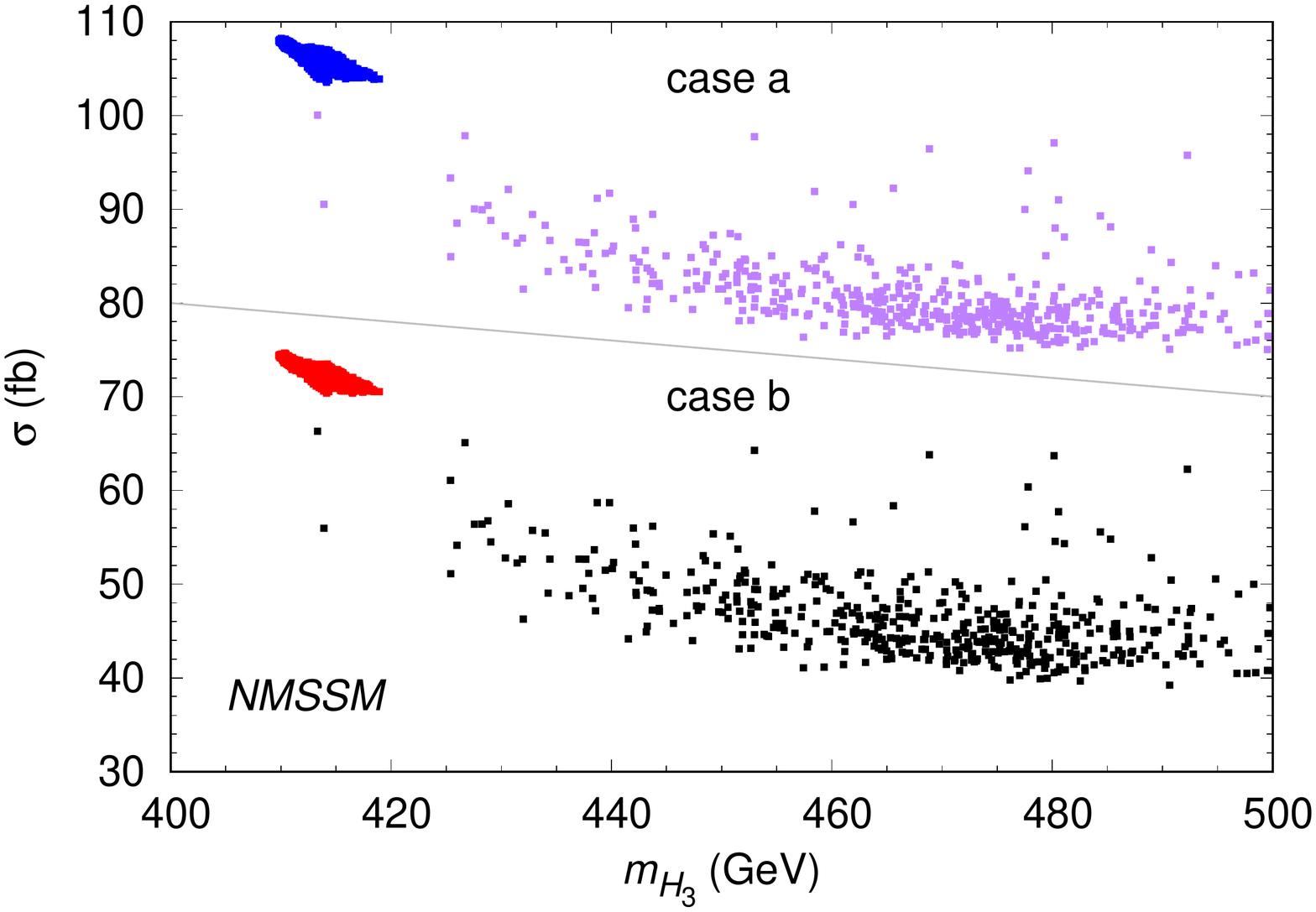} &
\hspace*{-1.8cm}\includegraphics*[width=10.5cm]{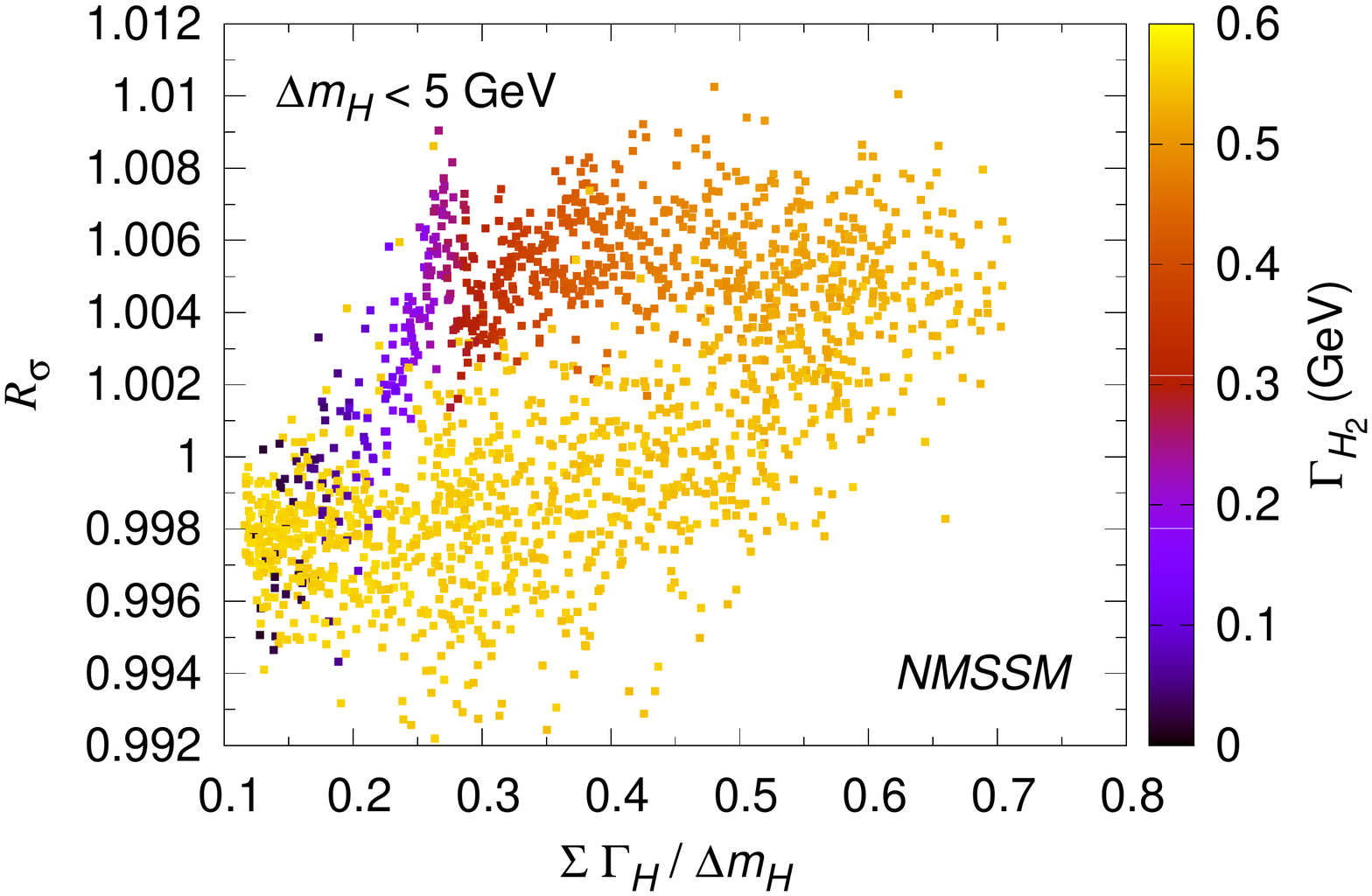} \\
\end{tabular}
\vspace*{-0.8cm}\caption{\label{fig:NM-sigma} Left -- Cross sections corresponding to the cases a (top half) and b (bottom half) for the scanned NMSSM points, with those shown in blue and red in the respective halves being the ones with $\Delta m_H<5$\,GeV. Right -- The ratio of the cross sections b and c as a function of the ratio of the sum of the widths of $H_2$ and $H_3$ and their mass difference, with the colour map showing the width of $H_2$, for the points with $\Delta m_H<5$\,GeV.}
\end{figure}

The top half of Fig.\,\ref{fig:NM-sigma} (left) shows the cross sections $\sigma_{\rm a}$ (top half) and $\sigma_{\rm b}$ (bottom half) with diagonal-only propagator matrix, as functions of the $H_3$ mass. One sees a large negative impact of the triangle-box interference, reducing the cross section uniformly by $\sim 35$\,fb for all the points. We note here that, in models with Supersymmetry, the box and triangle diagrams in principle include loops from squarks also. Here we take the view that the squarks are always too heavy to contribute significantly to either of these production processes (recall that we fixed the soft squark masses to 3\,TeV in our parameter space scans, to prevent the physical sparticle masses from conflicting with the direct search results from the LHC), and thus retain only the quark loops. A detailed study of the impact of the inclusion of squarks in the MSSM and the NMSSM (without the Higgs propagator interference effects) can be found in Refs.~\cite{Batell:2015koa,Huang:2017nnw,Huang:2019bcs}. The small blue and red islands near the lowest allowed $m_{H_3}$ and with overall largest cross sections in the top and bottom halves, respectively, are the points with $\Delta m_H < 5$\,GeV obtained from the secondary scan.

In the numerical calculation of the propagator matrix, in contrast, the (one-loop) Higgs self-energies due to all the relevant NMSSM particles were included.
The right panel of Fig.\,\ref{fig:NM-sigma}, however, shows negligible impact of introducing the full propagator matrix. This figure, restricted only to the points with $\Delta m_H <5$\,GeV, shows $R_\sigma$ against the ratio of the sum of $H_2$ and $H_3$ widths, $\sum \Gamma_H$, and $\Delta m_H$. Note that, for a more accurate picture, the widths used for producing this plot are the higher order ones output by {\tt NMSSMTools}, rather than the tree-level ones corresponding to the self-energies computed by our cross section code. $\sum \Gamma_H$ ranges between 535\,MeV and 565\,MeV for all the points, implying that when $\Gamma_{H_2}$, depicted by the colour map in the figure, reaches its maximum value, $\Gamma_{H_3}$ is at its minimum, and vice versa. The fact that the lowest $\Delta m_H$ obtained is 0.8\,GeV, according to Eq.\,(\ref{eq:masses-nmssm}), implies that   $\sum \Gamma_H/\Delta m_H$ is always smaller than 1 and hence the above mentioned condition of larger $\sum \Gamma_H$ than $\Delta m_H$ for a sizeable enhancement in the propagator effects is never met. Still, one can notice a small gradual increase in $R_\sigma$, meaning an increasing negative effect of the full propagator, as $\sum\Gamma_H$ rises with respect to $\Delta m_H$. This effect is  more pronounced for points with $H_2$ and $H_3$ widths closer to each other in magnitude, as illustrated by the violet/red points in the top left quadrant of the figure. A larger gap between these two widths, in contrast, generally tends to slightly increase $\sigma_{\rm c}$ compared to $\sigma_{\rm b}$ (the points in the bottom left quadrant).  

\begin{figure}[t!]
\begin{tabular}{ccc}
\hspace*{-1.3cm}\includegraphics*[width=8.2cm]{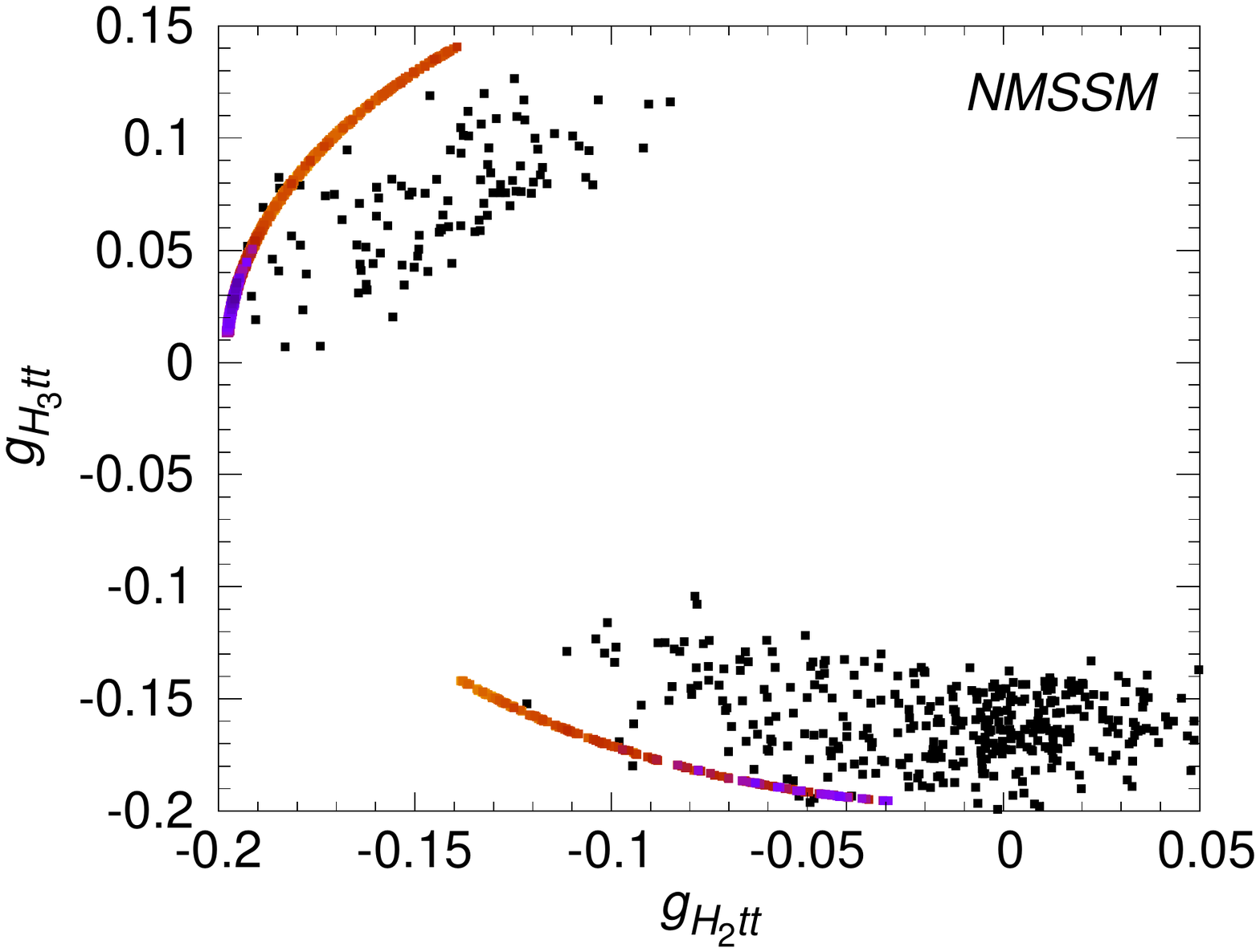} &
\hspace*{-3.5cm}\includegraphics*[width=8.2cm]{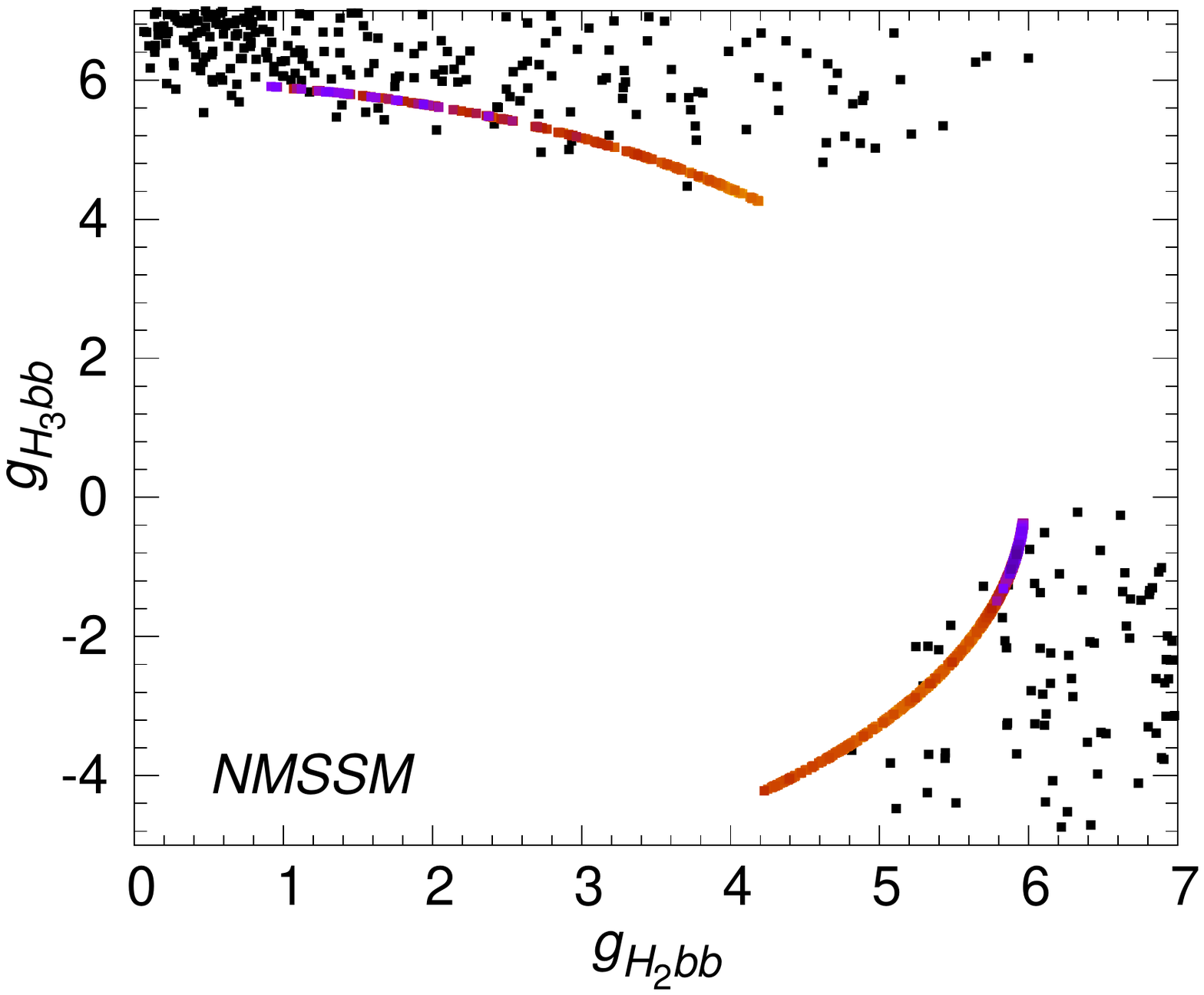} &
\hspace*{-3.2cm}\includegraphics*[width=8.2cm]{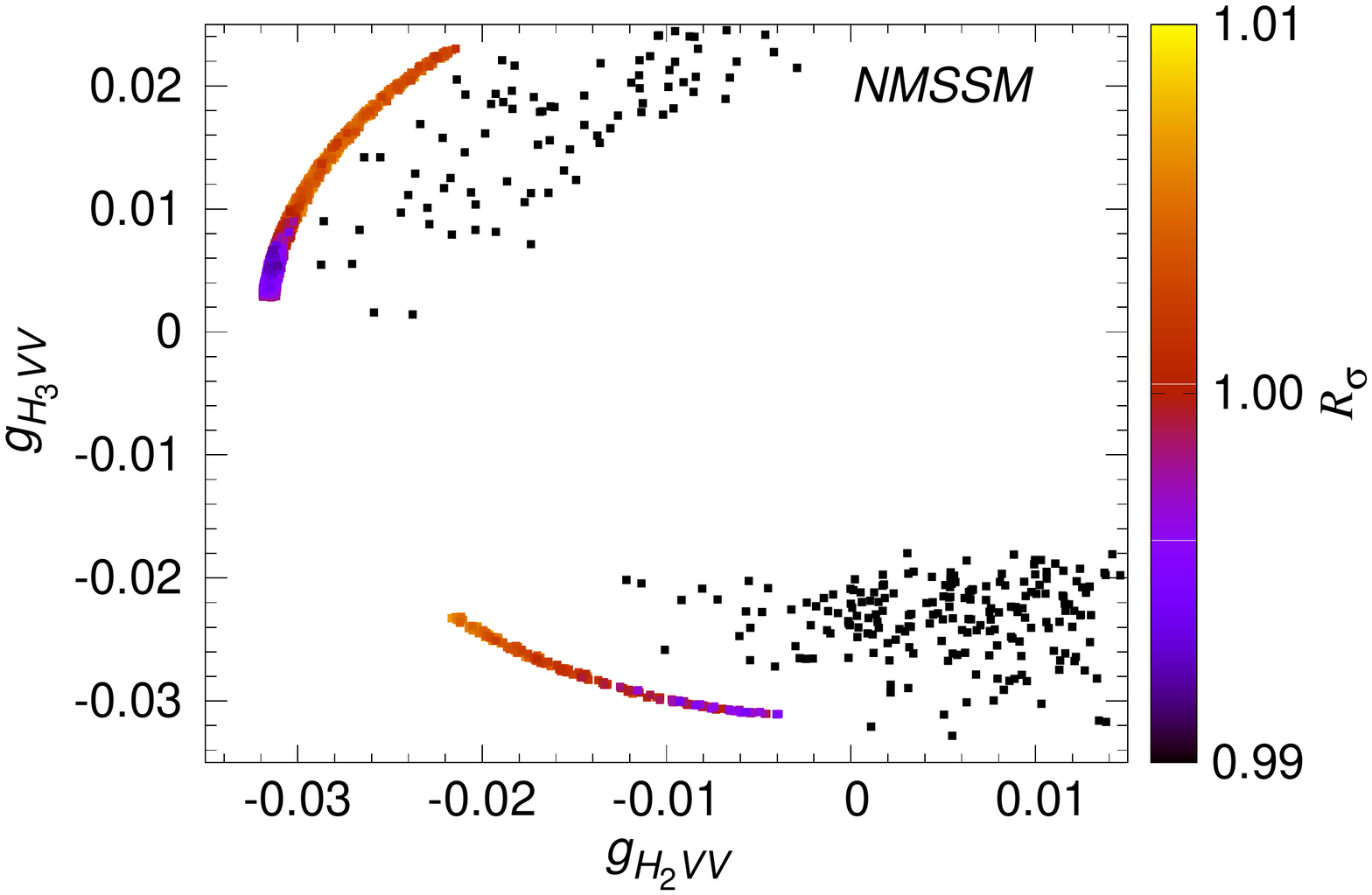} \\
\end{tabular}
\vspace*{-0.8cm}\caption{\label{fig:NM-params} $H_2$ and $H_3$ couplings to $t\bar{t}$ (left), $b\bar{b}$ (centre) and $VV$ (right) pairs in the NMSSM, with the colour map showing the ratio of the cross sections corresponding to cases b and c for the points with $\Delta m< 5$\,GeV.}
\end{figure}

The overall smallness of $R_\sigma$ in the NMSSM can be attributed partly to the large squark and slepton masses, so that their contribution to the Higgs self-energies is diminished, and partly to the specific Yukawa and gauge coupling combinations of the $H_2$ and $H_3$ in the narrow parameter space region yielding large mass degeneracy between these. Fig.\,\ref{fig:NM-params} shows $R_\sigma$ as a function of these coupling combinations. One notices in these figures that the colour-mapped points, which correspond to the parameter space region with $\Delta m_H <5$\,GeV, mark the boundaries of the (black) points from the extended scan. Thus, the search results from the LHC, besides directly constraining the mass of the $H_3$ to lie above $\sim 405$\,GeV, also restrict its top-Yukawa coupling to fairly small values, with either sign. The condition of mass degeneracy with $H_3$ then also dictates the signs and sizes of the $H_2$ couplings. 

According to the left panel of Fig.\,\ref{fig:NM-params}, while the $H_2$ and $H_3$ top-Yukawa couplings can take up three different sign combinations in general, for points with $\Delta m_H <5$\,GeV, the sign of $g_{H_2 t{\bar t}}$ is always negative, while that of $g_{H_3 t{\bar t}}$ can be both negative or positive. However, only positive $g_{H_3 t{\bar t}}$ values appear for large negative values of $g_{H_2 t{\bar t}}$. As the magnitude of the latter drops, that of the former increases, with $R_\sigma$ also rising slowly, until both reach equal values (with opposite signs). At that point, the sign of $g_{H_3 t{\bar t}}$ flips to negative, giving the largest $R_\sigma$ according to the colour map. A further increase in its magnitude, however, along with a decrease in the size of $g_{H_2 t{\bar t}}$, leads to a lowering of $R_\sigma$ again. In short, largest (allowed) values of one of the two top-Yukawa couplings, whether positive or negative, coupled with the smallest value of the other, results in $\sigma_{\rm c}>\sigma_{\rm b}$ and, as the two tend towards each other, $\sigma_{\rm c}$ starts to lower towards $\sigma_{\rm b}$ and eventually below it. 

The central panel of the figure likewise illustrates the impact of the variations in $g_{H_2 b{\bar b}}$ and $g_{H_3 b{\bar b}}$ on $R_\sigma$. Note that the points in the bottom half of this plot correspond to the points in the top half of the left panel, and vice versa. Thus, the sign of the bottom-Yukawa coupling of a given Higgs boson is always opposite to that of its top-Yukawa coupling, so that $g_{H_2 b{\bar b}}$ is positive only, conversely to $g_{H_2 t{\bar t}}$. Furthermore, $R_\sigma$  shows a similar trend with the variation in the sizes of $g_{H_2 b{\bar b}}$ and $g_{H_3 b{\bar b}}$ as with the top-Yukawa couplings -- the largest (allowed) value of one bottom-Yukawa coupling paired with the smallest value of the other yields $\sigma_{\rm c}>\sigma_{\rm b}$, while $\sigma_{\rm c} \leq \sigma_{\rm b}$ results from their comparable magnitudes. The dependence of $R_\sigma$ on the relative signs and magnitudes of $g_{H_2 VV}$ and $g_{H_3 VV}$ follows the behaviour of the top-Yukawa couplings exactly, as seen in the right panel of the Fig.\,\ref{fig:NM-params}. Their allowed values are, however, much smaller than even those of the top-Yukawa couplings, pointing towards the decoupling regime of the (N)MSSM. As for the remaining couplings of the $H_2$ and $H_3$, even when the corresponding (s)particles have sufficiently low masses, including $A_1$ as well as $\chi_{1,2}^0$ and $\chi_{1}^\pm$ (which are higgsino-like and thus have masses $\sim \mu_{\rm eff}\sim 150$\,GeV, see Eq.\,(\ref{eq:masses-nmssm})), their influence on $R_\sigma$ is too small to merit a discussion here.


\subsection{The Type-II N2HDM}
\label{subsec:num-n2hdm}

\begin{figure}[t!]
\begin{tabular}{cc}
\hspace*{-1.0cm}\includegraphics*[width=9.5cm]{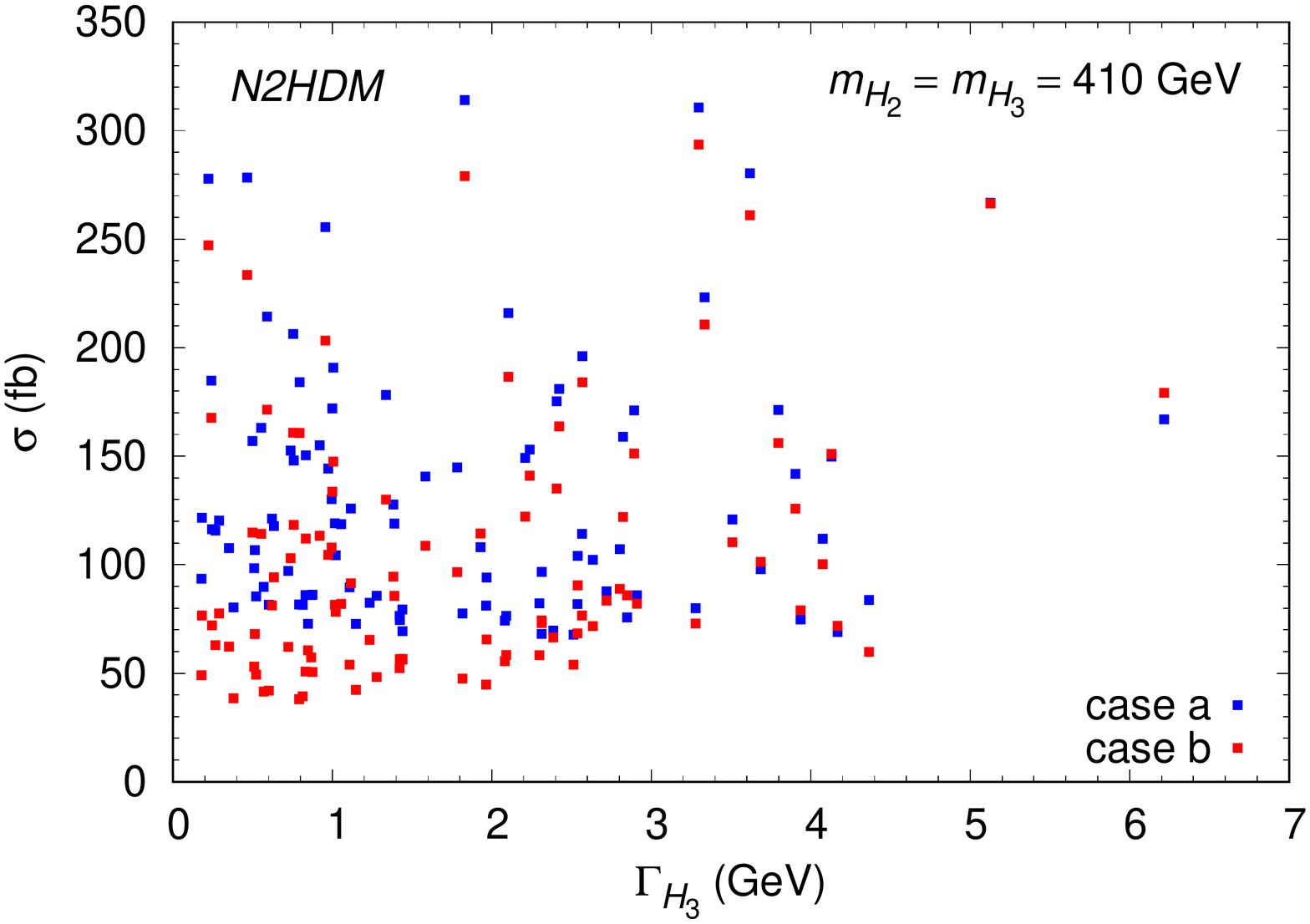} &
\hspace*{-1.2cm}\includegraphics*[width=9.5cm]{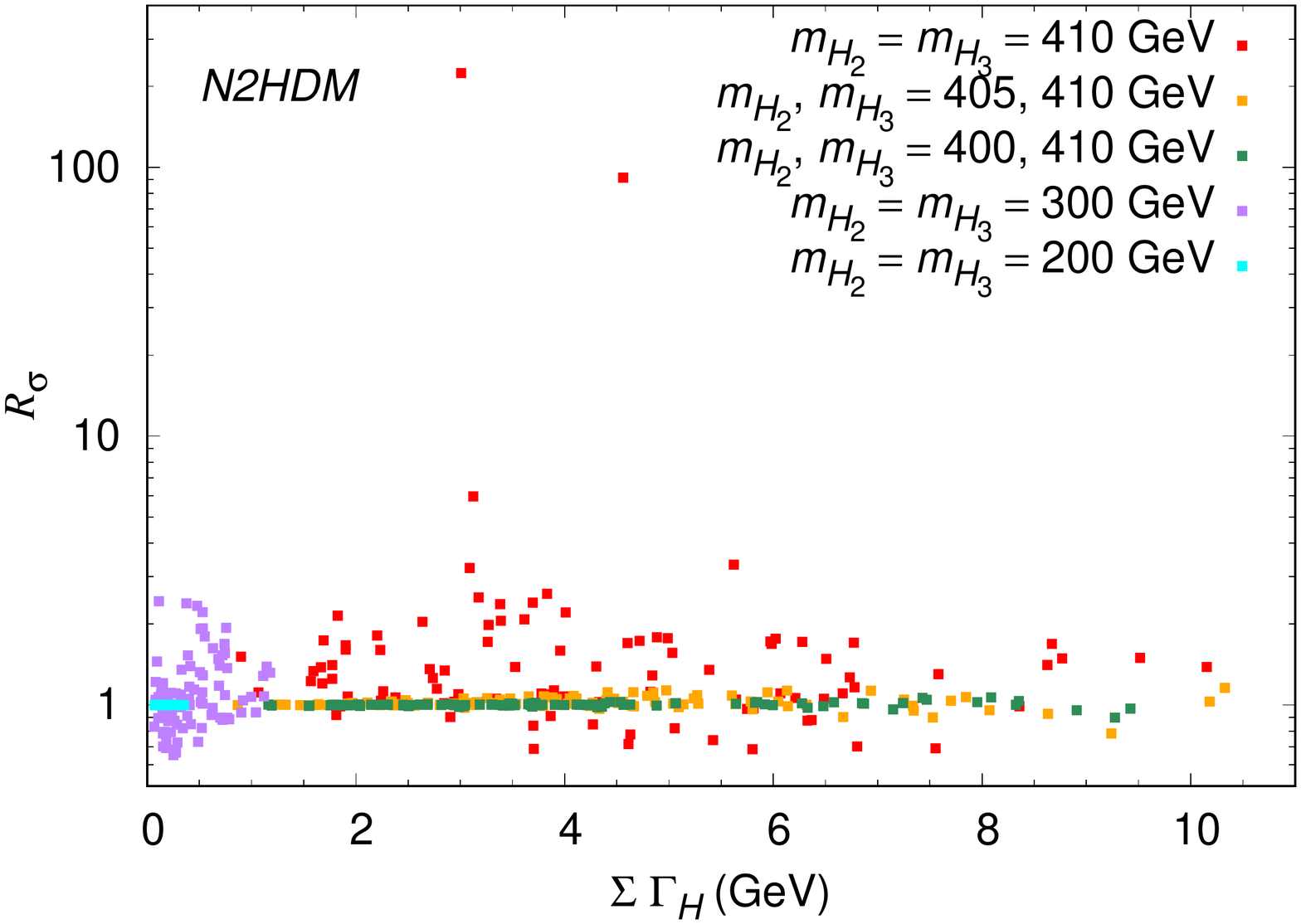} \\
\end{tabular}
\vspace*{-0.8cm}\caption{\label{fig:N2-sigma} Left -- Cross sections for the points obtained from the N2HDM parameter space scan with $m_{H_2}=m_{H_2}=410$\,GeV, corresponding to the cases a (blue) and b (red) as functions of $\Gamma_{H_3}$. Right -- The ratio of the cross sections b and c against the sum of the widths of $H_2$ and $H_3$ for the points obtained from the five N2HDM scans with different $m_{H_2}$ and $m_{H_3}$ configurations.}
\end{figure}

As indicated earlier, the Higgs boson masses are input parameters in the N2HDM, which allows us to investigate $H_2$ and $H_3$ with exactly equal masses, that can also be much lower than those obtained in the NMSSM. For a direct comparison with the NMSSM though, in our first scan for this model we set $m_{H_2}=m_{H_3}=410$\,GeV, and the $\sigma_{\rm a}$ and $\sigma_{\rm b}$ for the 100 successful points thus obtained are shown in blue and red, respectively, in the left panel of Fig.\,\ref{fig:N2-sigma} against the width of $H_3$. Contrary to the NMSSM, triangle-box interference does not reduce the cross section uniformly for all the points. While for most of the points $\sigma_b$ is smaller by a few tens of fb than $\sigma_a$, the former is larger than the latter by upto 10\,fb for a few points. This is owing to the wider ranges of magnitudes as well as sign combinations for the $H_2$ and $H_3$ couplings being available in this model, as will be explained later. Notice also that $\Gamma_{H_2}$ can reach a few GeVs and, in fact, $\Gamma_{H_3}$ can simultaneously be quite large, as illustrated by the horizontal axis of the right panel of the figure. Once again, $\Gamma_{H_{2,3}}$ here are the widths output by {\tt ScannerS}, instead of the tree-level ones that can be obtained from the one-loop self-energies computed by our code. The vertical axis of this panel shows the impact of including the full propagator matrix in the cross section calculation. 

For a number of points from the first scan with $m_{H_2}=m_{H_3}=410$\,GeV, seen in red in  Fig.\,\ref{fig:N2-sigma} (right), $\sigma_{\rm b}$ is a few times larger than $\sigma_{\rm c}$, but for some points it gets reduced by upto 35\%, implying a net positive contribution from the off-diagonal terms in the matrix. For two of the red points, though, $R_{\sigma}$ exceeds 100, meaning a two orders of magnitude reduction in  $\sigma_{\rm b}$. (We point out here that these two points were omitted from the left panel to keep the y-axis scale visually interesting, but the corresponding cross sections will be provided below.) To assess the effect of reducing the mass degeneracy, the points from scans with $\Delta m_H=5$\,GeV and $\Delta m_H=10$\,GeV, with $m_{H_3}$ still fixed to 410\,GeV, are also plotted in this figure in orange and green, respectively. Evidently, a larger $\Delta m_H$ results in smaller fluctuations in $\sigma_{\rm b}$, as the $R_\sigma$ value lies very close to 1 for all the 100 green points. The violet points in the figure correspond to the scan with $m_{H_2}=m_{H_3}=300$\,GeV. While in general $R_\sigma$ can deviate substantially from 1 for these points also, its maximum value does not exceed 3. The main reason for this is that the widths of $H_2$ and $H_3$ are always lower than 1\,GeV in this case, unlike the $m_{H_2}=m_{H_3}=410$\,GeV case, owing to the fact that their masses lie below the $t\bar{t}$ production threshold. Lowering $m_{H_2}$ and $m_{H_3}$ even further to under the $H_1H_1$ threshold expectedly results in a vanishing impact of the off-diagonal propagator matrix terms, as demonstrated by the cyan points in the figure, which are all clustered together below $\sum\Gamma_H \lesssim 200$\,MeV.

\begin{table}
\centering\begin{tabular}{|c|cccccc|} \hline
  Parameter/Observable  & BP1      & BP2      & BP3      & BP4      & BP5      & BP6   \\ 
                            \hline \hline
  $m_A$\,(GeV)          & 712.2  & 772.67  & 640.04  & 601.21 & 658.33 & 630.11 \\
  $m_{H^\pm}$\,(GeV)    & 709.04  & 776.41  & 654.53  & 604.04  & 663.11 & 654.45 \\
  $m_{12}^2$\,(GeV$^2$) & 84725.4  & 71277.6  & 82115.1  & 61133.1 & 69580.1 & 65586.7 \\
  $\tan\beta$           & 1.3     & 1.0    & 1.3    & 2.0   & 1.8   & 1.2   \\
  $g_{H_1t{\bar t}}$    & 1.024    & 1.038    & 0.955    & 0.981   & 0.989   & 0.986   \\
  $g_{H_1VV}$           & 1.000      & 1.000      & 0.954    & 0.990    & 1.000     & 0.930    \\
  sign(${\cal R}_{13}$)         & $-$      & $+$      & $-$      & $+$     & $-$     & $+$     \\
  ${\cal R}_{23}$              & $-$0.671 & $-$0.569 & $-$0.921 & 0.887   & 0.436   & 0.870    \\
  $v_S$\,(GeV)          & 1511.3  & 2357.5  & 1945.8   & 1667.5 & 2025.9 & 2459.4 \\
 \hline
  \hline
 $g_{H_2t{\bar t}}$          & 0.545    & 0.766    & $-$0.092 & 0.106    & 0.533    & $-$0.089 \\
 $g_{H_3t{\bar t}}$          & 0.505    & 0.509    & 0.805    & $-$0.533 & $-$0.203 & $-$0.827 \\ \hline
 $g_{H_1b{\bar b}}$          & 0.959    & 0.956    & 0.952    & 1.024    & 1.030     & 0.846  \\
 $g_{H_2b{\bar b}}$          & $-$0.984 & $-$0.879 & $-$0.636 & $-$0.998 & $-$1.490   & $-$0.771 \\
 $g_{H_3b{\bar b}}$          & $-$0.880  & $-$0.627 & $-$1.202 & 1.684    & 0.831    & 1.091  \\ \hline
 $g_{H_2VV}$                 & $-$0.029 & $-$0.024 & $-$0.289 & $-$0.120  & 0.038    & $-$0.362  \\
 $g_{H_3VV}$                 & $-$0.143 & $-$0.037 & 0.077    & $-$0.079 & 0.038    & $-$0.061  \\ \hline
 $g_{H_1AA}$                 & 82.120    & 111.190   & 56.667   & 51.450    & 66.070    & 61.080     \\
 $g_{H_2AA}$                 & $-$2.585 & $-$3.303 & $-$22.422& $-$9.804 & 2.731    & $-$30.780  \\
 $g_{H_3AA}$                 & $-$1.142 & $-$4.892 & 6.175    & $-$3.003 & 3.429    & $-$3.474  \\ \hline
 $g_{H_1AZ}$, $g_{H_1H^+W^-}$ & $-$0.031 & $-$0.041 & $-$1.398 & 0.018    & 0.017    & $-$0.069  \\
 $g_{H_2AZ}$, $g_{H_2H^+W^-}$ & $-$0.741 & $-$0.822 & $-$0.262 & $-$0.446 & $-$0.899 & $-$0.334  \\
 $g_{H_3AZ}$, $g_{H_3H^+W^-}$ & $-$0.671 & $-$0.568 & $-$0.965 & 0.895    & $-$0.437 & 0.940      \\ \hline
 $g_{H_1H^+H^-}$              & 81.039   & 112.584  & 60.978   & 52.263   & 67.588   & 68.084    \\
 $g_{H_2H^+H^-}$              & $-$2.554 & $-$3.336 & $-$23.728& $-$9.903 & 2.788    & $-$33.502 \\
 $g_{H_3H^+H^-}$              & $-$1.127 & $-$4.943 & 6.523    & $-$3.068 & 3.487    & $-$3.930   \\ \hline
 $g_{H_1H_1H_1}$             & 3.006    & 3.319    & 4.774    & 7.359    & 7.472    & 0.159     \\
 $g_{H_1H_1H_2}$             & $-$1.512 & $-$1.260  & $-$6.579 & $-$2.557 & 1.404    & $-$7.562  \\
 $g_{H_1H_1H_3}$             & $-$1.083 & $-$1.351 & 1.709    & $-$2.238 & 0.666    & $-$0.206 \\
 $g_{H_1H_2H_2}$             & 0.314    & 5.460     & 2.100      & 2.411    & 1.883    & 7.209   \\
 $g_{H_1H_2H_3}$             & 0.325    & 3.736    & $-$0.231 & $-$1.453 & $-$0.954 & $-$1.479 \\
 $g_{H_1H_3H_3}$             & 0.226    & 2.682    & 1.249    & 4.826    & 0.296    & 8.859    \\
 $g_{H_2H_2H_2}$             & $-$63.753& $-$46.913& $-$30.048& $-$44.402& $-$95.745& $-$22.313\\
 $g_{H_2H_2H_3}$             & 3.092    & 0.163    & $-$0.700   & 6.525    & 1.830     & 7.870   \\
 $g_{H_2H_3H_3}$             & $-$3.725 & $-$3.740  & $-$4.920  & $-$2.443 & 1.860     & $-$8.990 \\
 $g_{H_3H_3H_3}$             & $-$50.644& $-$27.989& $-$74.038& 98.871   & 53.169   & 55.571   \\ \hline
 \hline
 $\Gamma_{H_2}$\,(GeV)       & 1.63 & 3.13  & 3.15  & 0.58   & 1.58   & 4.67 \\
 $\Gamma_{H_3}$\,(GeV)       & 1.38 & 1.43  & 3.62   & 1.78   & 0.27   & 3.68   \\ 
 \hline
 \hline
  $\sigma_{2\times 2}$\,(fb) & 122.9 & 99.2  & 102.9 & 204.9 & 93.0  & 120.7 \\
  $\sigma_{\rm a}$\,(fb)     & 35518.6 & 13465.4 & 280.4 & 144.9 & 115.8 & 98.0  \\
  $\sigma_{\rm b}$\,(fb)     & 34536.1 & 13417.6 & 260.1   & 96.6   & 62.9  & 101.3  \\
  $\sigma_{\rm c}$\,(fb)     & 154.3 & 146.7  & 153.1 & 96.2  & 63.6  & 102.6 \\ \hline
\end{tabular}
\caption{\label{tab:N2-BPs} Values of the input parameters, couplings and widths of the Higgs bosons, together with the cross sections corresponding to the six selected BPs of the N2HDM.}
\end{table}

For a detailed investigation, we selected six benchmark points (BPs) from our main scan with $m_{H_2}=m_{H_3}=410$\,GeV. The input parameters, the widths and couplings of $H_2$ and $H_3$ as well as the four cross sections corresponding to these points are given in Table\,\ref{tab:N2-BPs}. $\sigma_{2\times 2}$ in the table implies the cross section obtained by setting $m_{H_3}\to\infty$ (in order to decouple the $H_3$), with all the other input parameters fixed to their exact values for a given BP, and is quoted for reference. BP1 and BP2 are the two points with the highest $R_\sigma$ in Fig.\,\ref{fig:N2-sigma}, for BP3 and BP4 the $R_\sigma$ value lies very close to 1 while BP5 and BP6 are chosen from amongst the points for which $\sigma_c$ is slightly enhanced compared to $\sigma_b$.

In the 2HDM, and the N2HDM by extension, of the type-II, the $B$-physics measurement strongly constrain $m_{H^\pm}$~\cite{Muhlleitner:2016mzt,Arbey:2017gmh} and therefore the latter lies above 600\,GeV for all the successful points from the scans, while $\tan\beta$ is also pushed to smaller values, as can be noted in the table. $m_A$ is then also restricted to values close to $m_{H^\pm}$ by the EW precision constraints. One feature distinguishing the points with the largest $R_\sigma$ (BP1 and BP2) from the rest of the BPs are the larger $m_{H^\pm}$ and $g_{H_1 t\bar{t}}$ values and relatively small $\tan\beta$. Such parameter configurations result in specific combinations of the couplings of $H_2$ and $H_3$ for BP1 and BP2, which in turn lead to very high $R_\sigma$ for these. For these two points, $g_{H_2 t\bar{t}}$ and $g_{H_3 t\bar{t}}$ are both positive and large while $g_{H_2 b\bar{b}}$, $g_{H_3 b\bar{b}}$, $g_{H_2VV}$ and $g_{H_3VV}$ are all negative. In the case of BP3, $g_{H_2 t\bar{t}}$ and $g_{H_3 t\bar{t}}$ have signs opposite to each other while $g_{H_2 b\bar{b}}$ and $g_{H_3 b\bar{b}}$ are both negative. We note here that, again in contrast with the NMSSM, $g_{H_2 b\bar{b}}$ is negative for all the 100 N2HDM points for the $m_{H_2}=m_{H_3}=410$\,GeV scenario, and also that, for a majority of these points, three out of the four Yukawa couplings had the same signs. BP4 and BP5 are very similar points in that the two top-Yukawa couplings have signs that are opposite not only to each other but also to the signs of the corresponding bottom-Yukawa couplings. BP6 is the only point of its kind found in the scan, with $g_{H_2 t\bar{t}}$, $g_{H_3 t\bar{t}}$ and $g_{H_2 b\bar{b}}$ all having negative signs, and it therefore uniquely exhibits a constructive triangle-box interference as well as constructive propagator interference, so that $\sigma_{\rm a}<\sigma_{\rm b}<\sigma_{\rm c}$.

\begin{figure}[tbp]
\centering\begin{tabular}{cc}
\vspace*{-0.4cm}\hspace*{-0.6cm}$g_{H_2t\bar t}$ & \hspace*{-1.4cm}$g_{H_2b\bar b}$ \\
\vspace*{-0.8cm}\hspace*{-0.8cm}\includegraphics*[width=8.3cm]{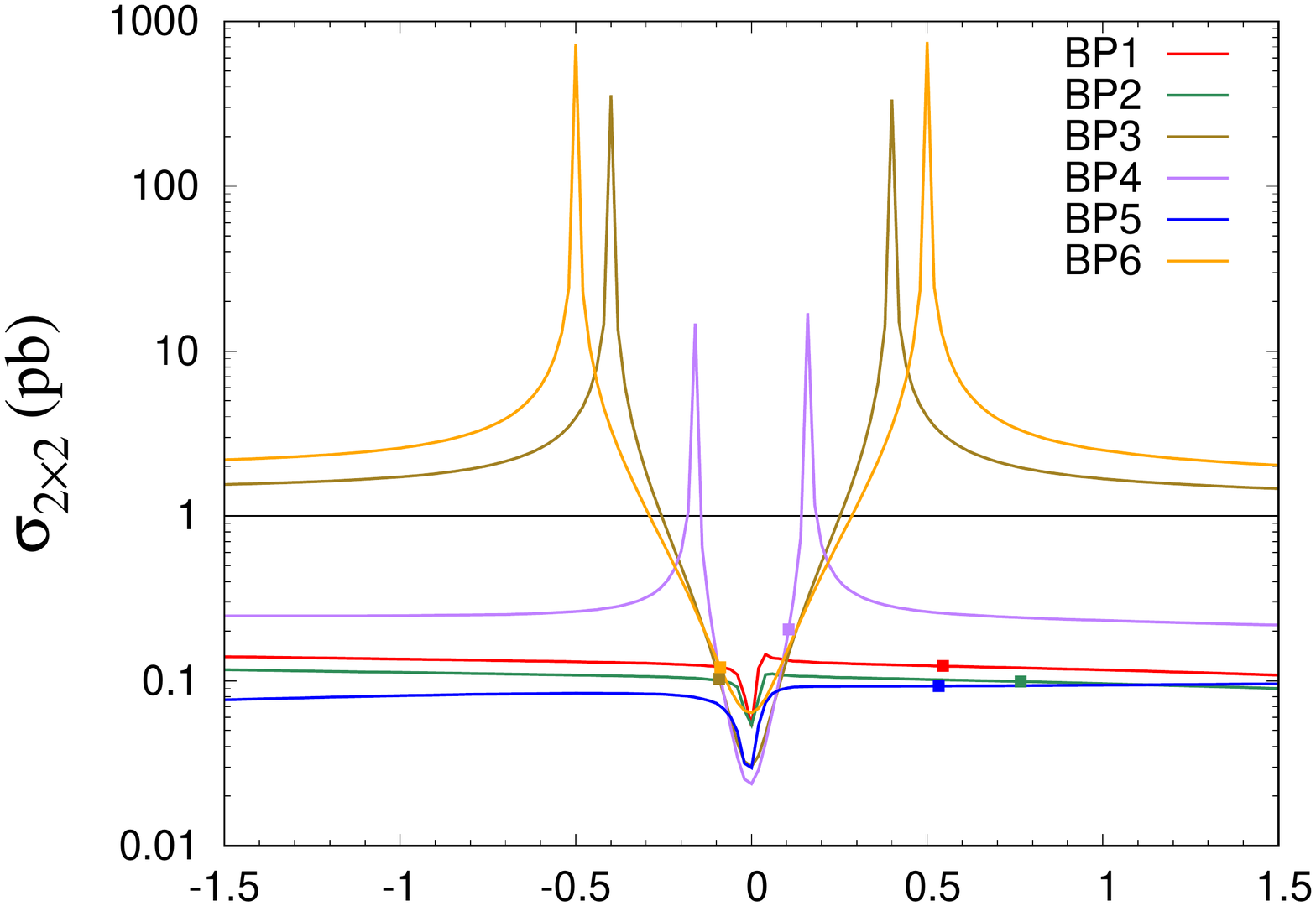} & \hspace*{-1.6cm}\includegraphics*[width=8.3cm]{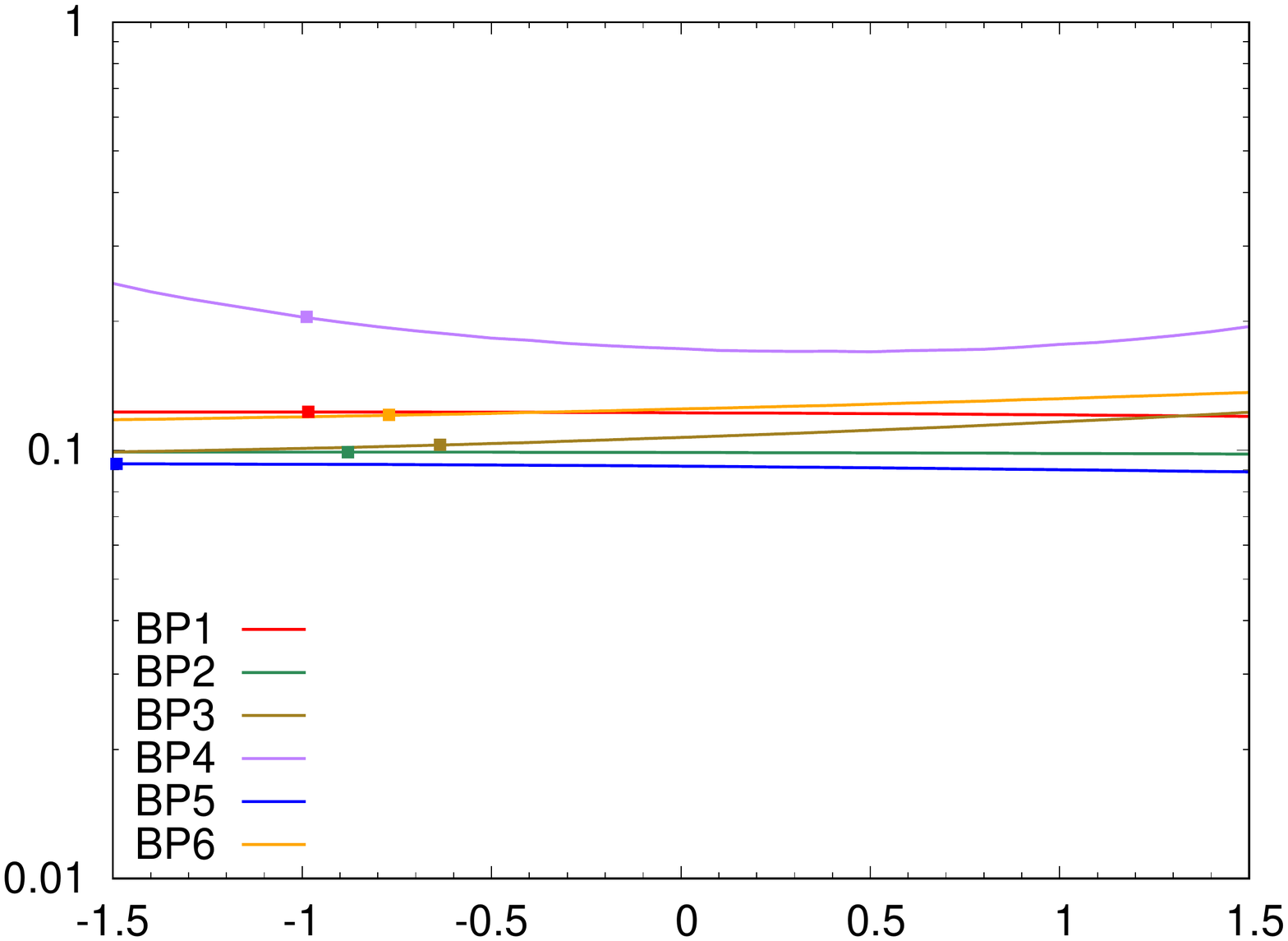} \\
\vspace*{-0.8cm}\hspace*{-0.8cm}\includegraphics*[width=8.3cm]{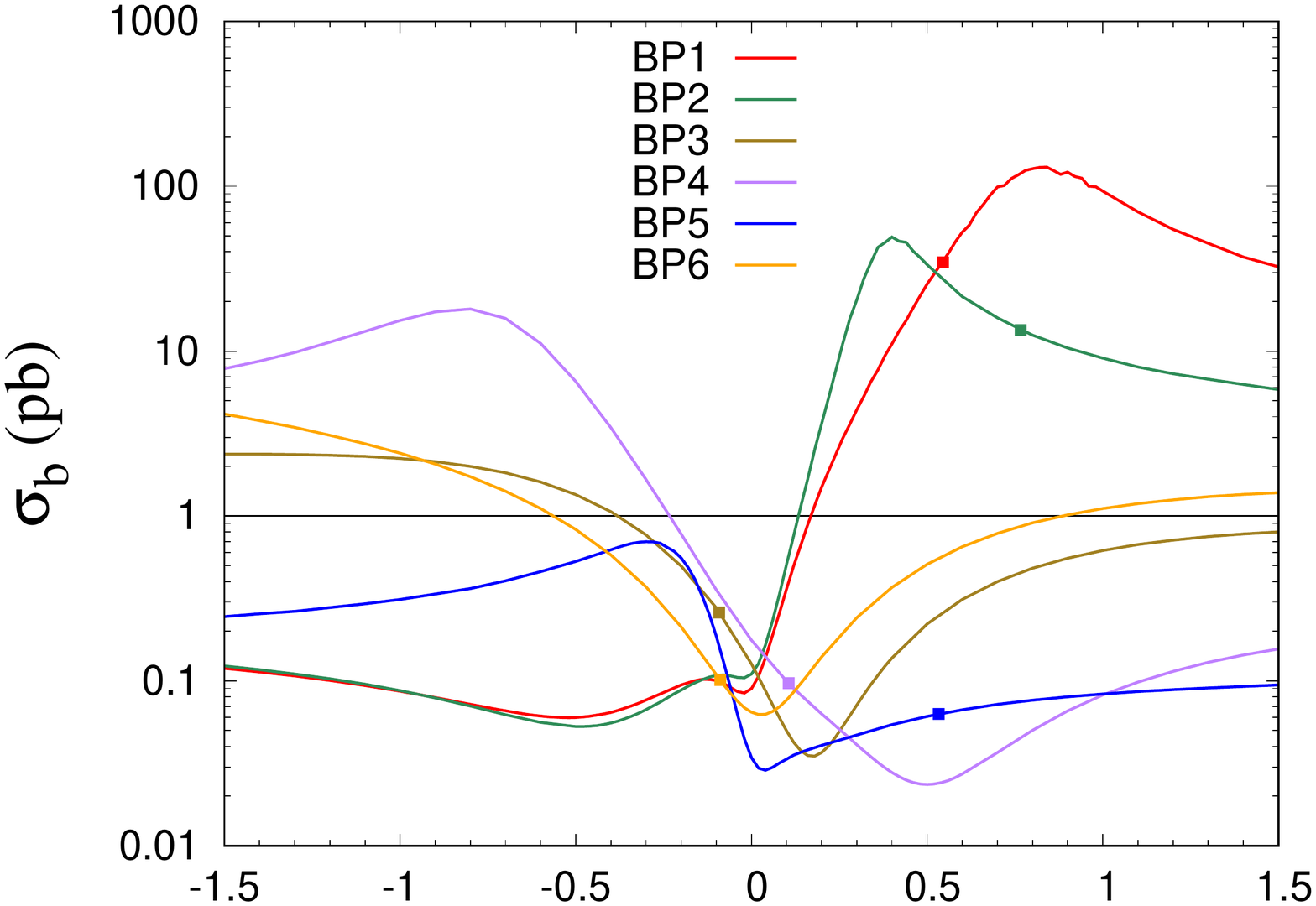} & 
\hspace*{-1.6cm}\includegraphics*[width=8.3cm]{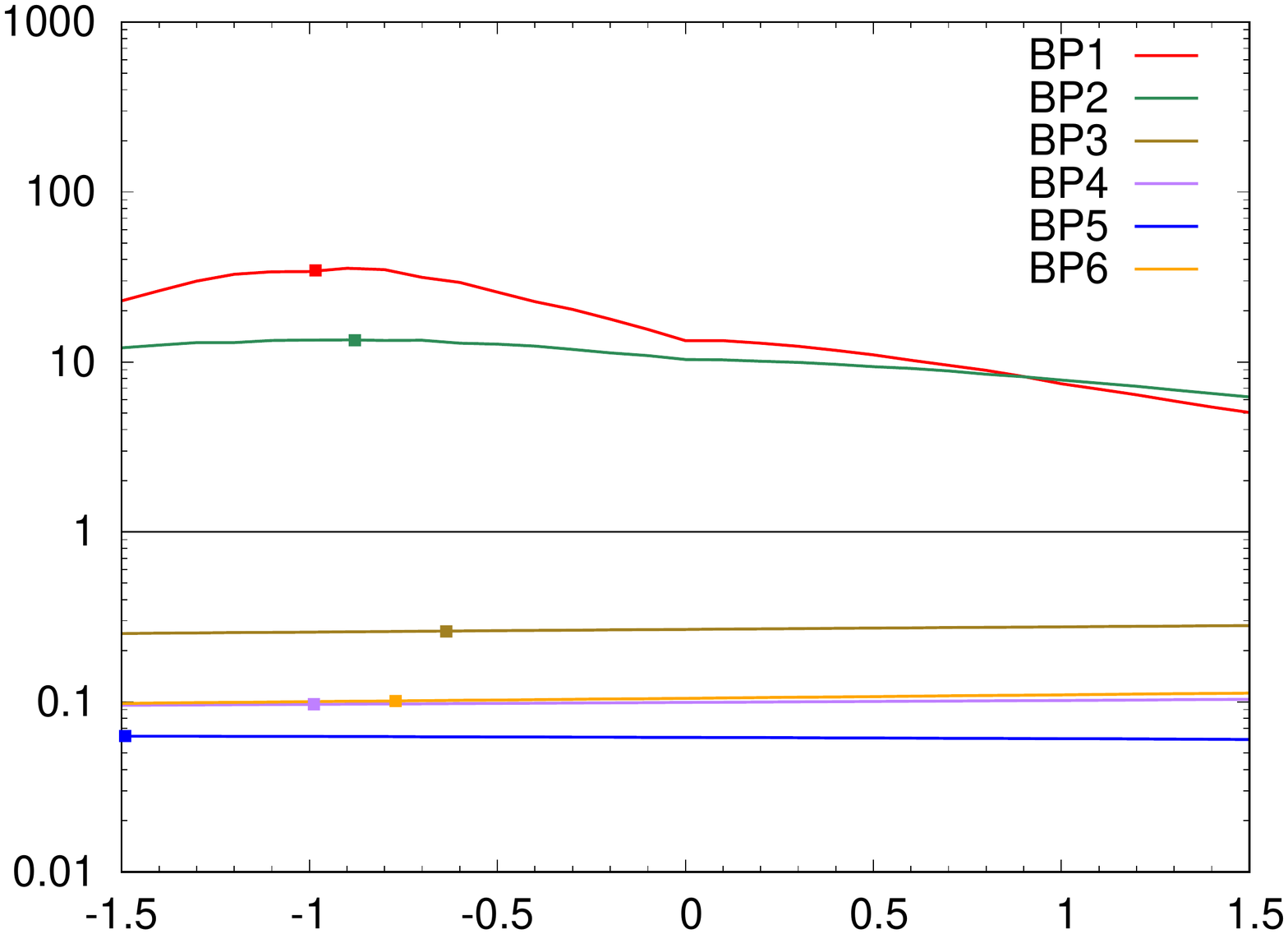} \\
\vspace*{-0.8cm}\hspace*{-0.8cm}\includegraphics*[width=8.3cm]{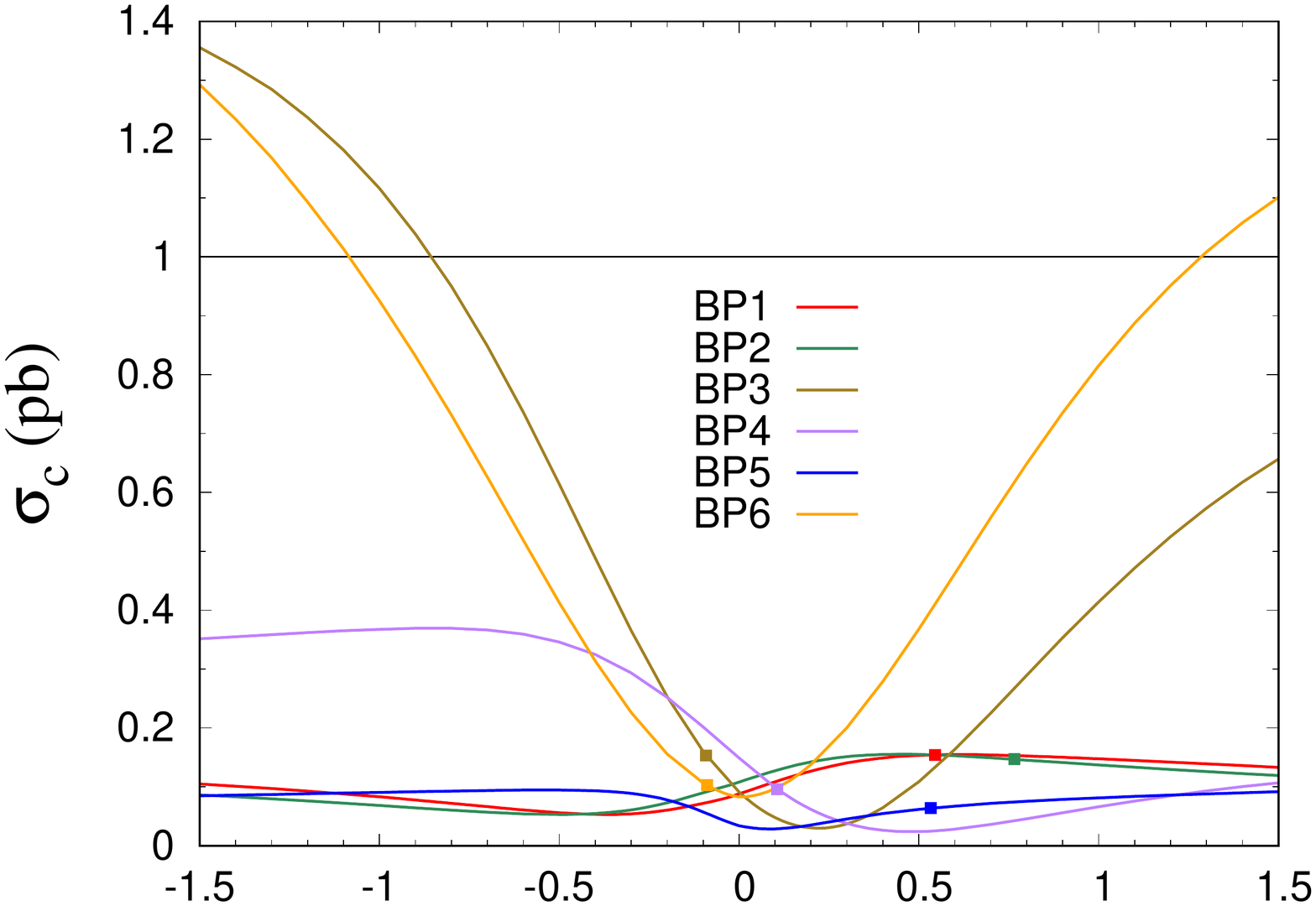} &
\hspace*{-1.6cm}\includegraphics*[width=8.3cm]{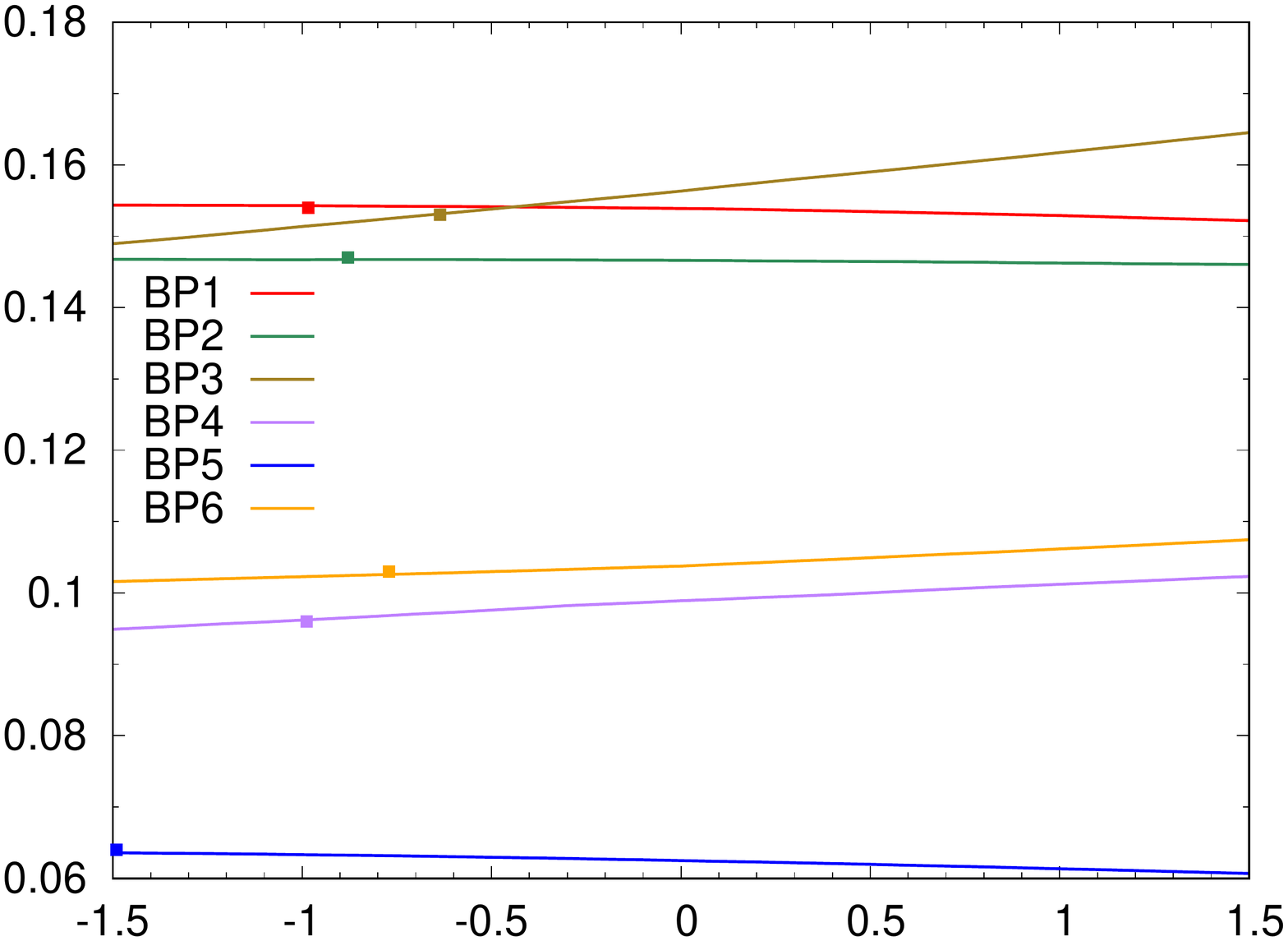} \\
\hspace*{-0.8cm}\includegraphics*[width=8.3cm]{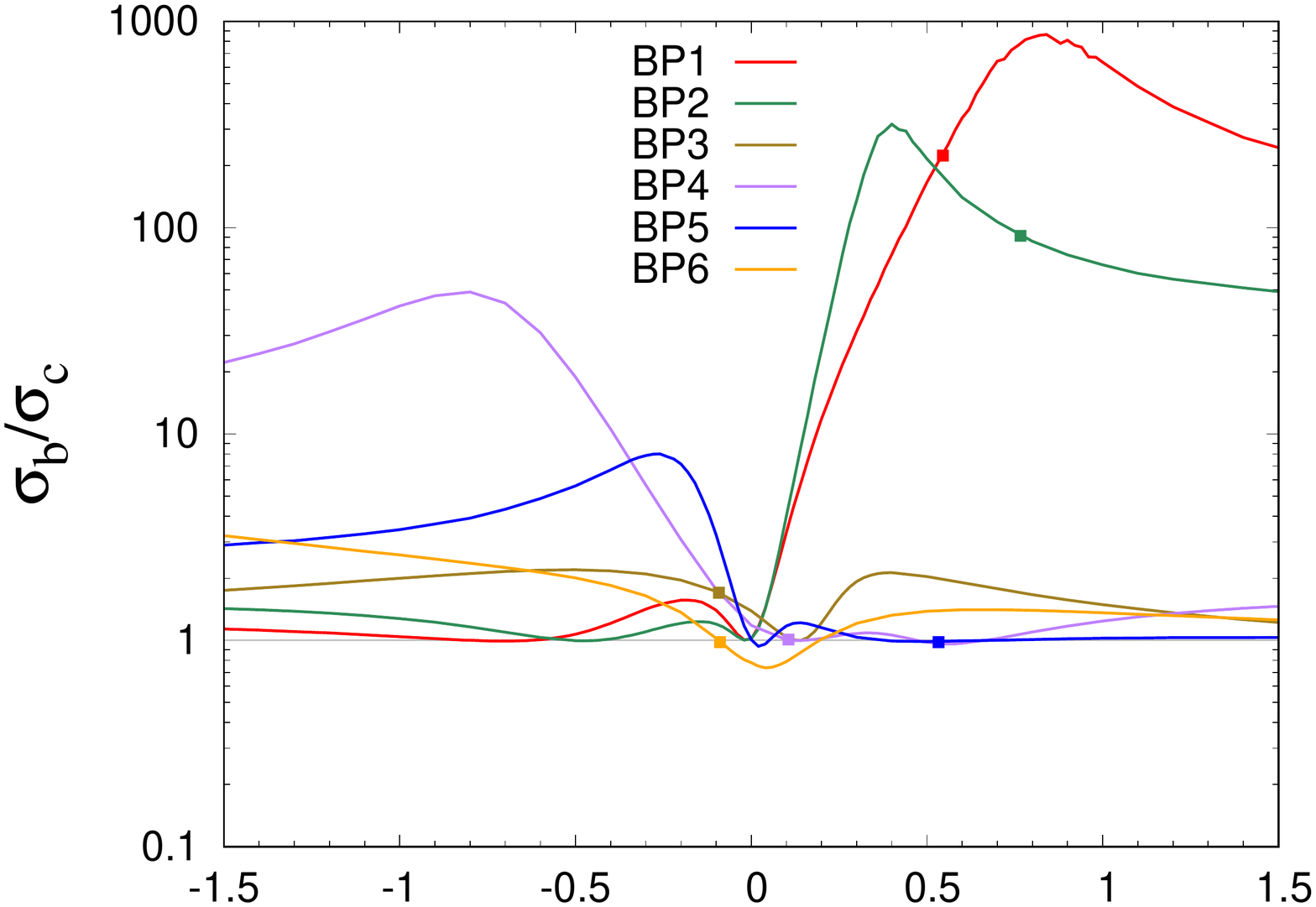} &
\hspace*{-1.6cm}\includegraphics*[width=8.3cm]{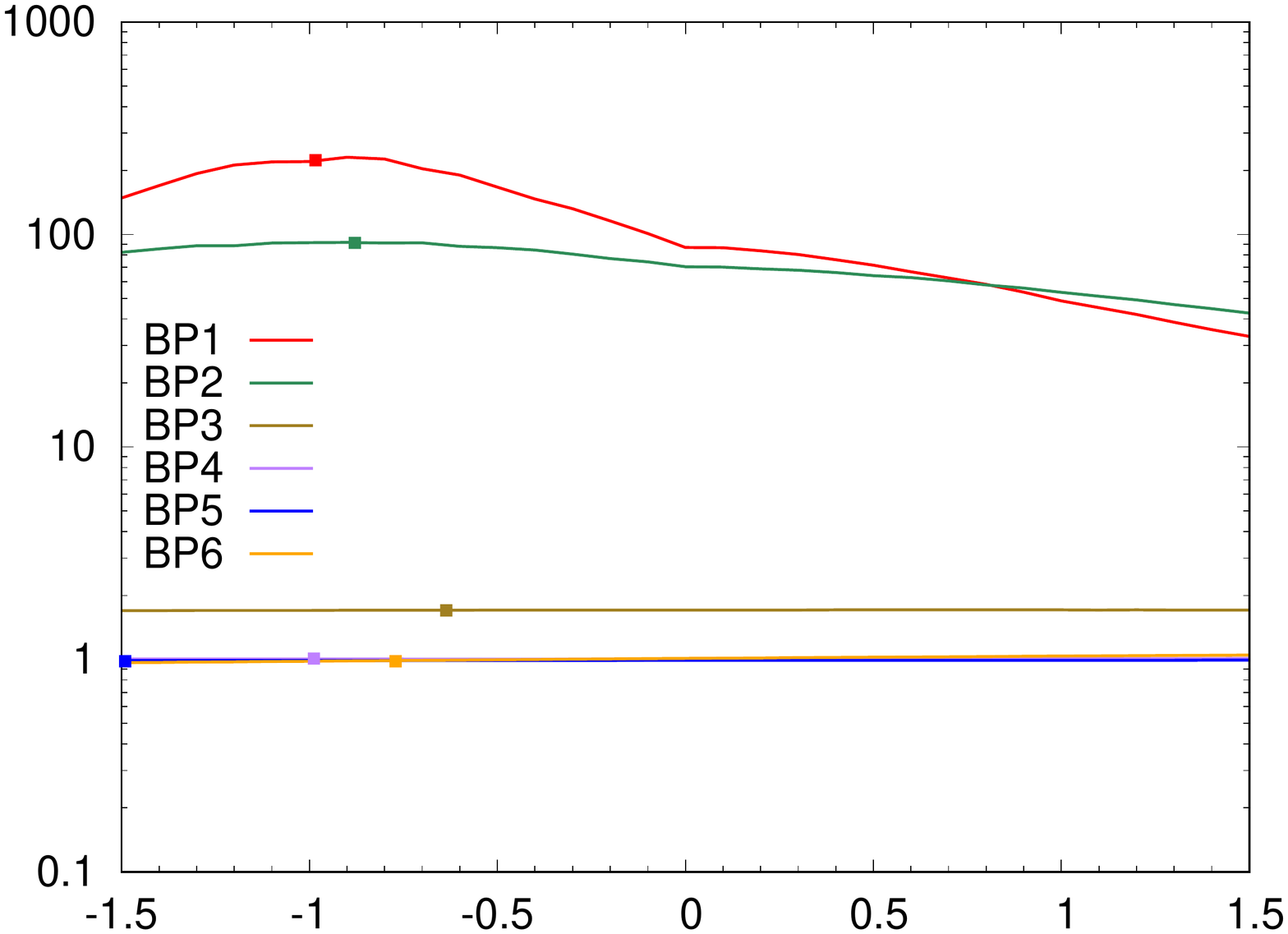} \\
\end{tabular}
\vspace*{-0.5cm}\caption{\label{fig:N2-params-1} Various cross sections as functions of the couplings $g_{H_2t\bar t}$ (left) and $g_{H_2b\bar b}$ (right). The point on a line marks the actual value of the plotted coupling for the corresponding BP. See text for more details.}
\end{figure}

\begin{figure}[tbp]
\centering\begin{tabular}{cc}
\vspace*{-0.4cm}\hspace*{-0.6cm}$g_{H_2VV}$ & \hspace*{-1.4cm}$g_{H_2H_1H_1}$ \\
\vspace*{-0.8cm}\hspace*{-0.8cm}\includegraphics*[width=8.3cm]{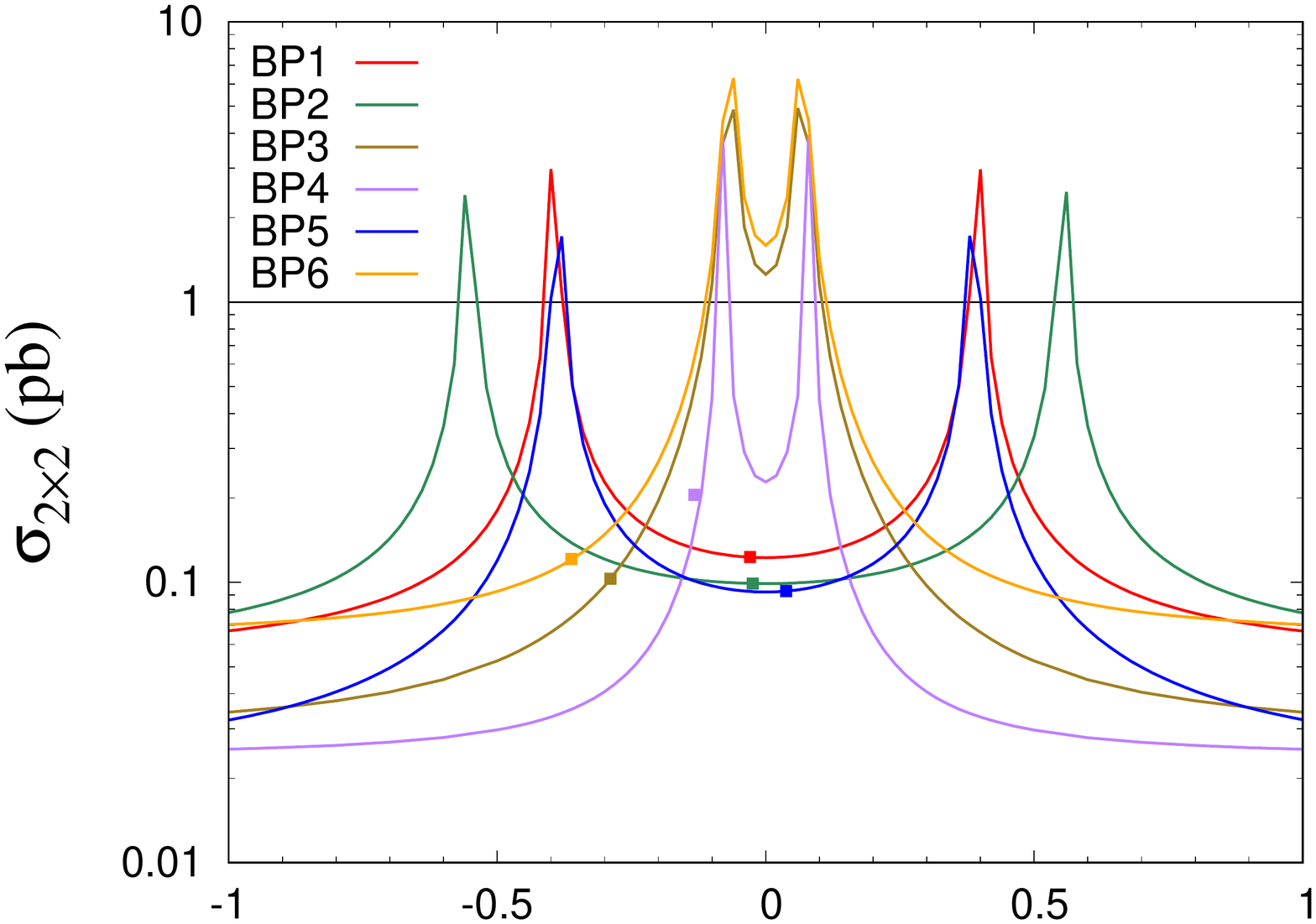} & \hspace*{-1.6cm}\includegraphics*[width=8.3cm]{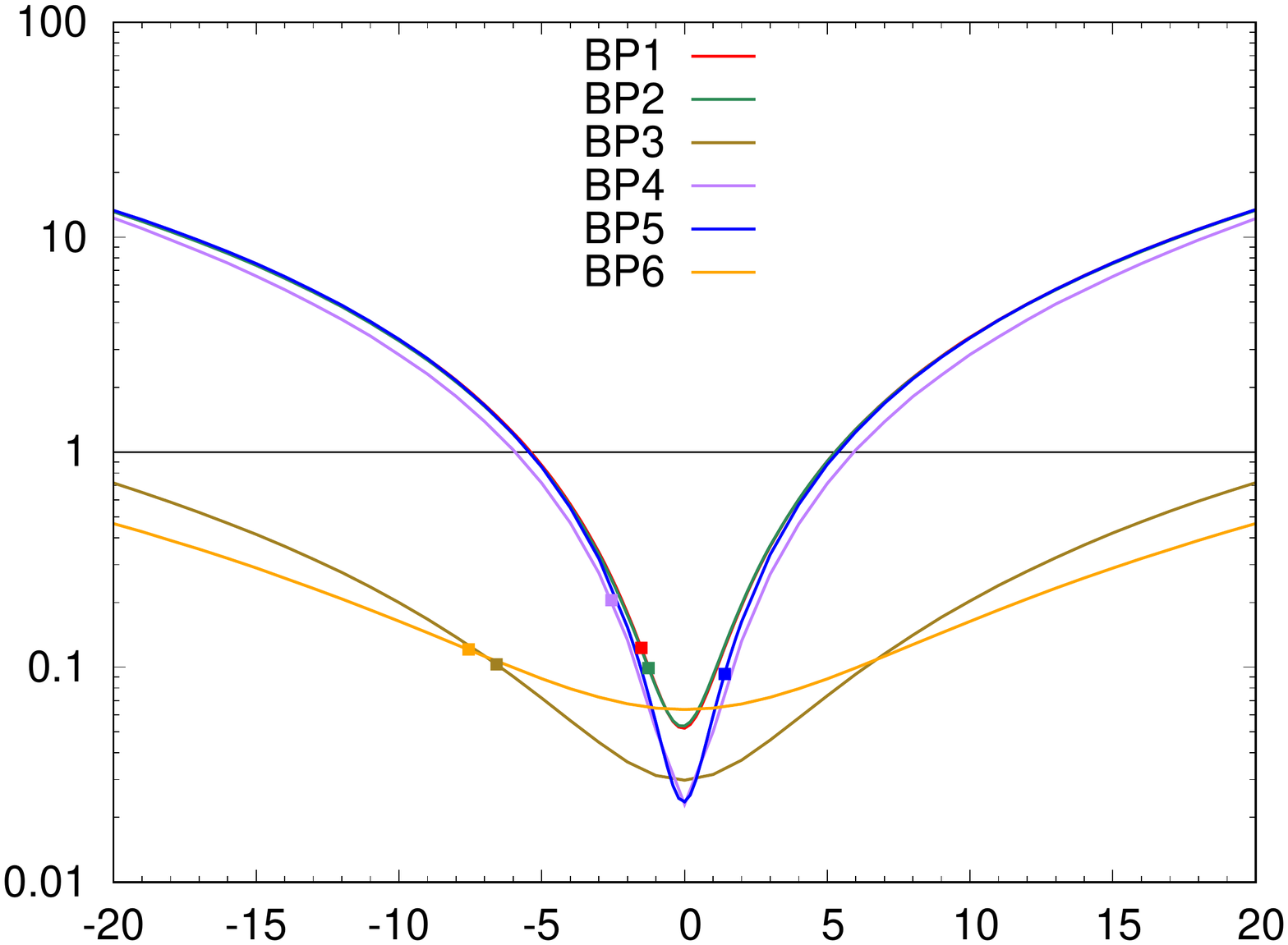} \\
\vspace*{-0.8cm}\hspace*{-0.8cm}\includegraphics*[width=8.3cm]{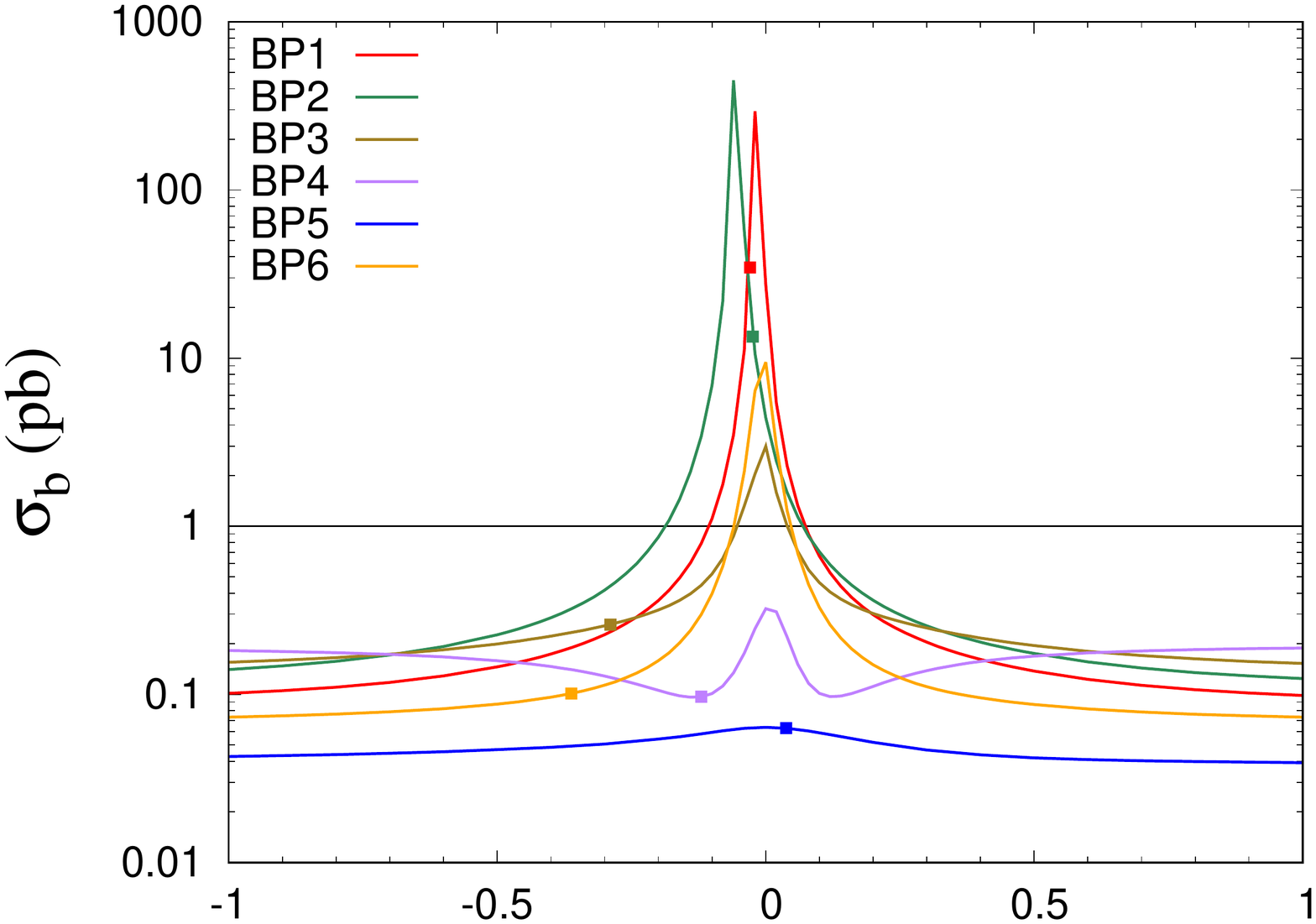} & 
\hspace*{-1.6cm}\includegraphics*[width=8.3cm]{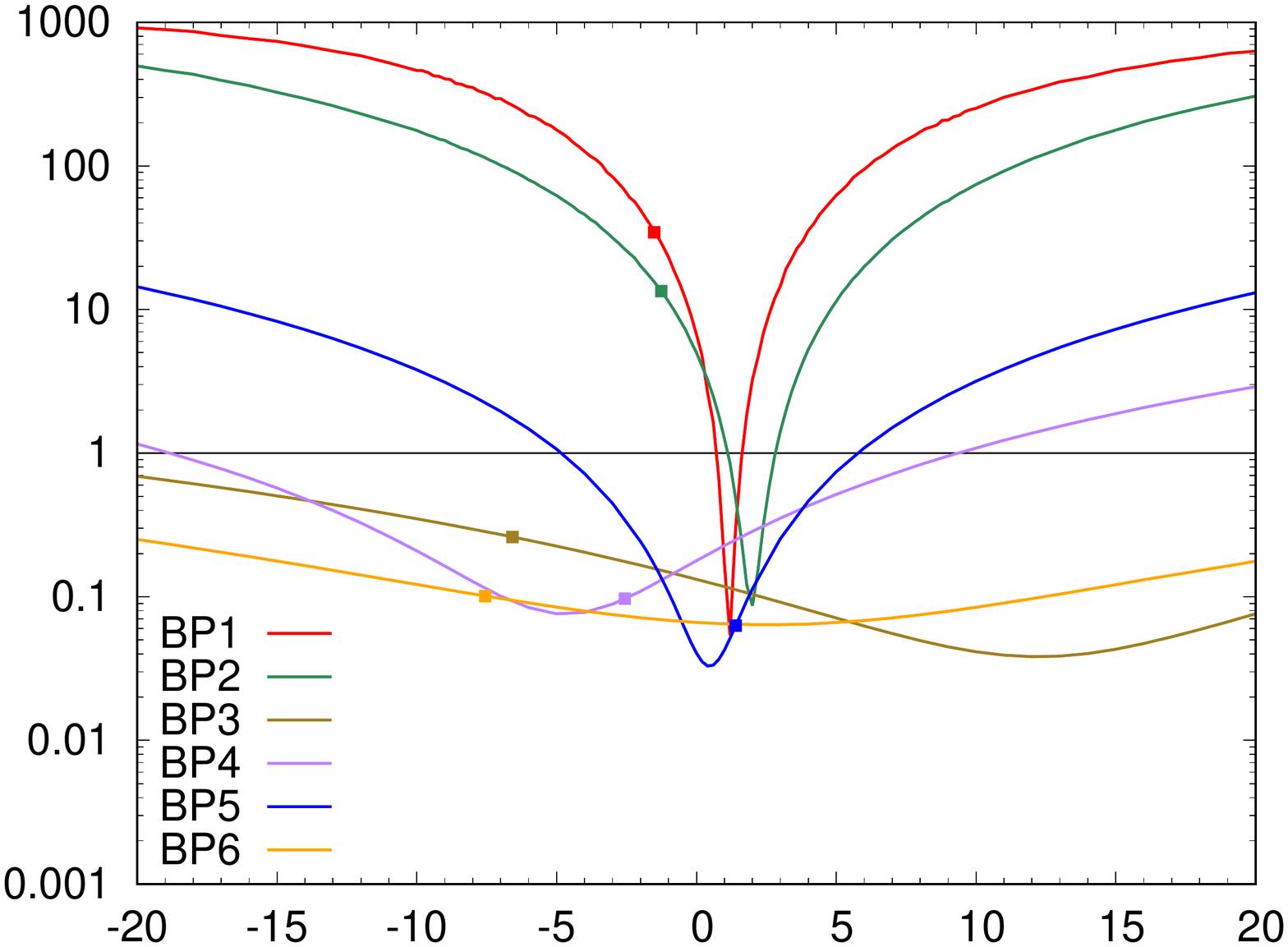} \\
\vspace*{-0.8cm}\hspace*{-0.8cm}\includegraphics*[width=8.3cm]{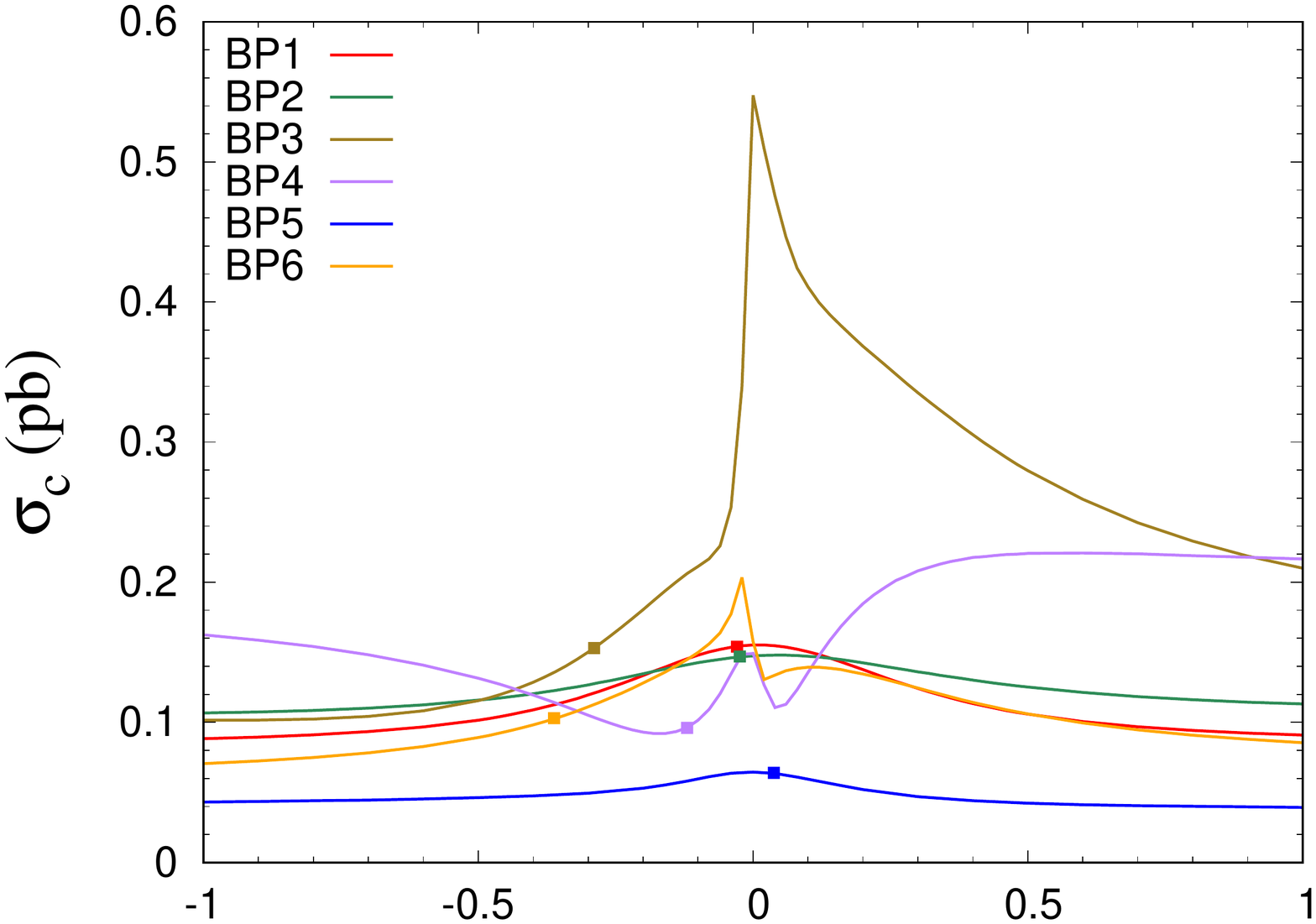} &
\hspace*{-1.6cm}\includegraphics*[width=8.3cm]{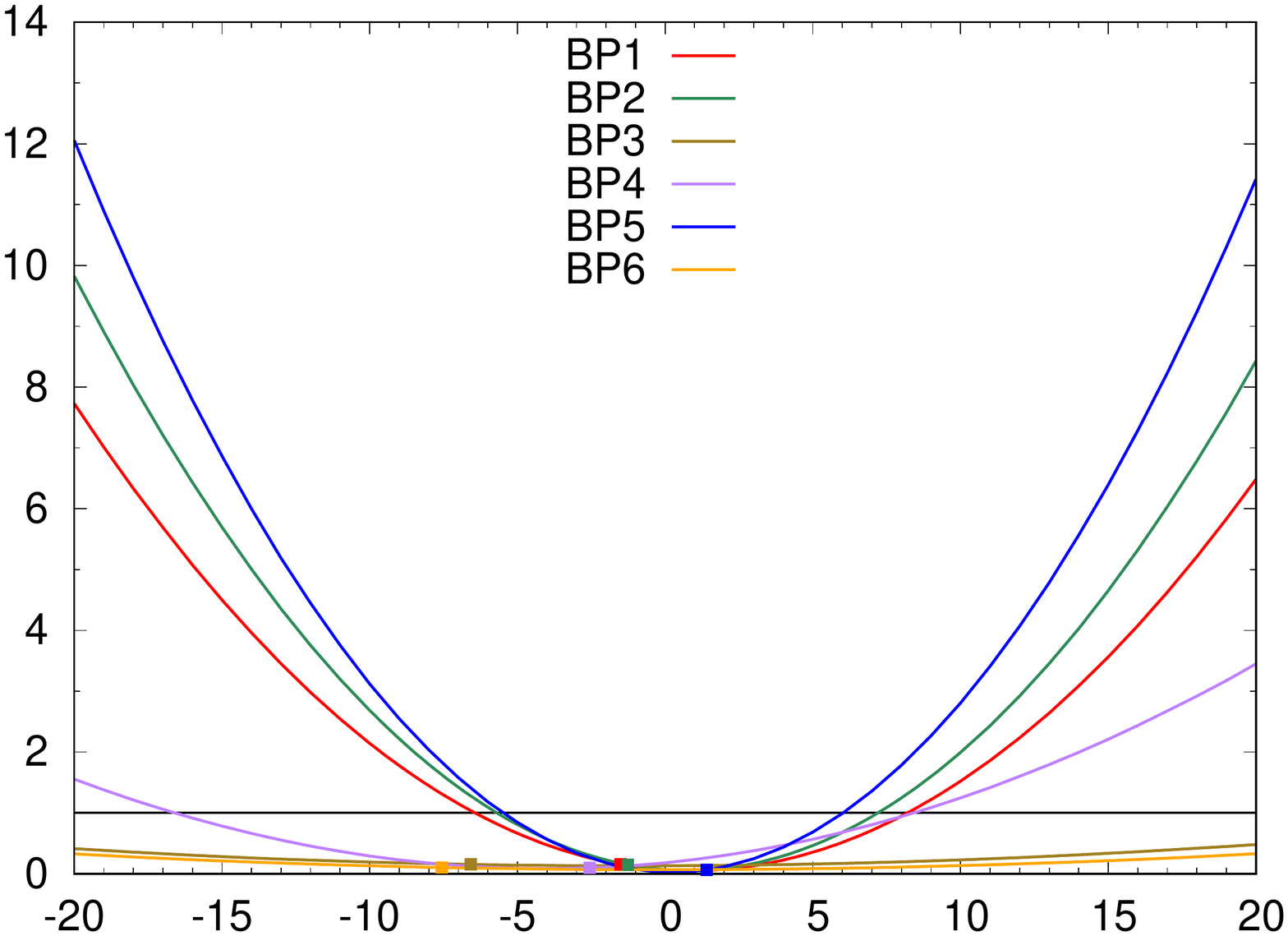} \\
\hspace*{-0.8cm}\includegraphics*[width=8.3cm]{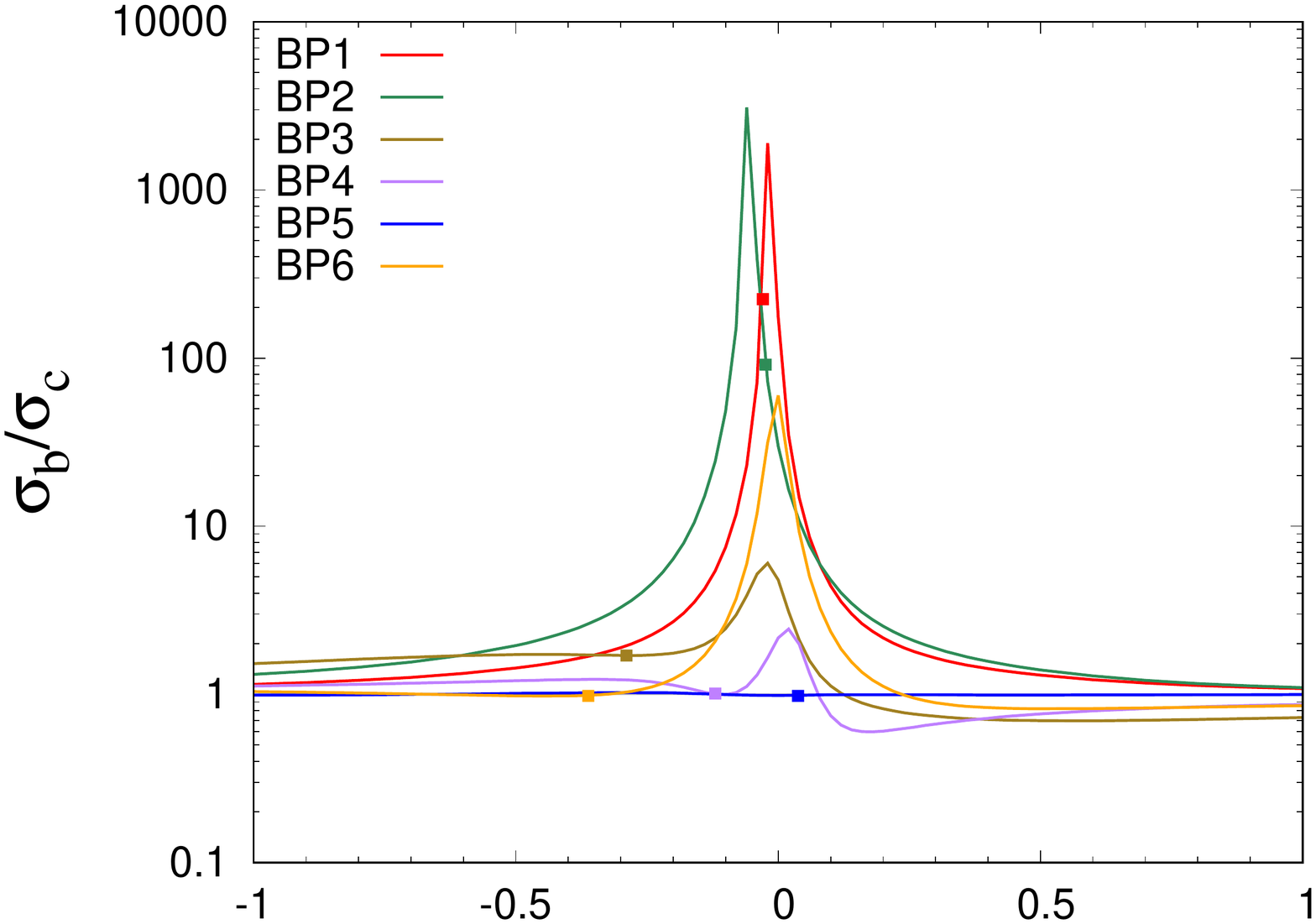} &
\hspace*{-1.6cm}\includegraphics*[width=8.3cm]{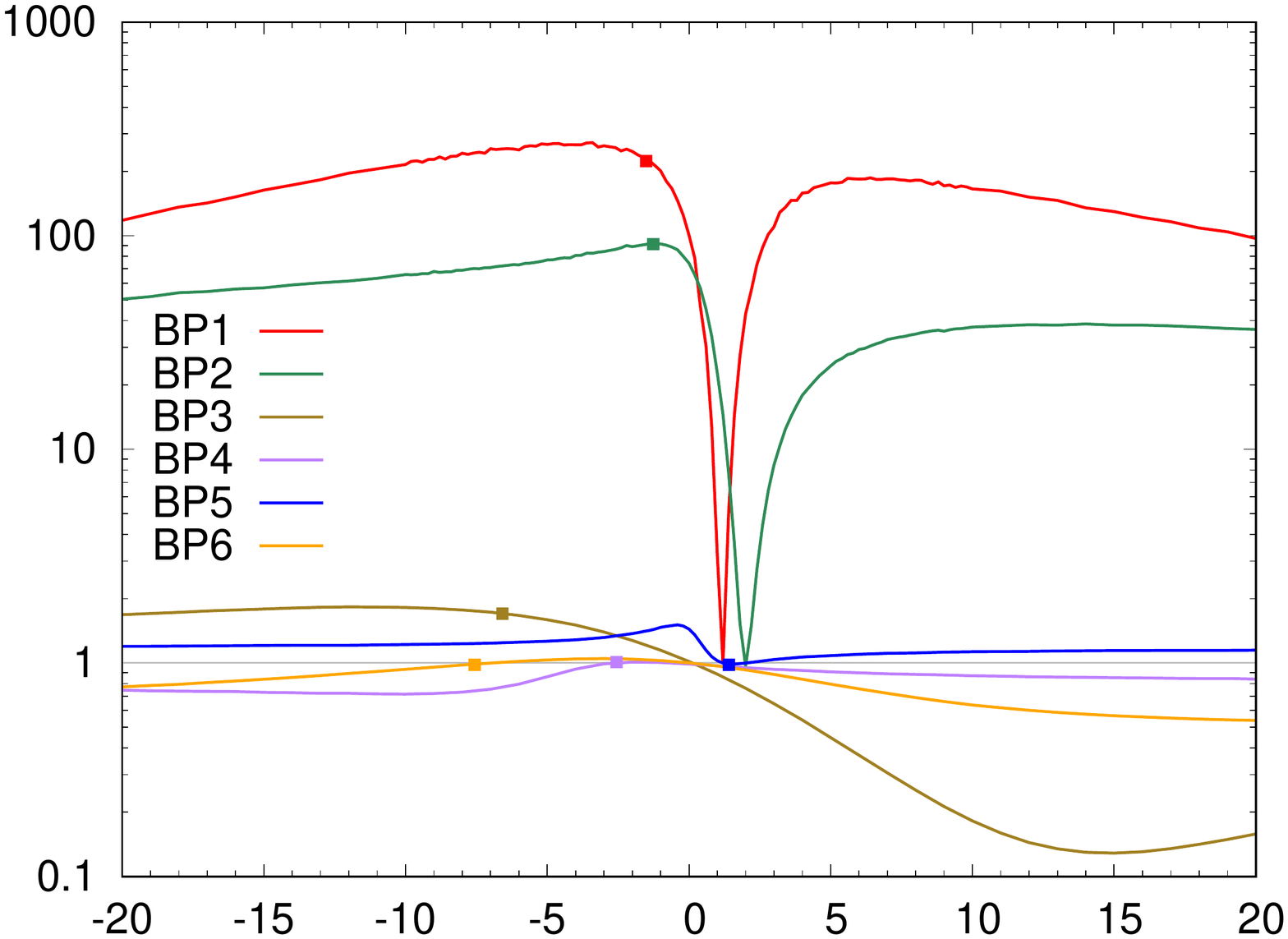} \\
\end{tabular}
\vspace*{-0.5cm}\caption{\label{fig:N2-params-2} Various cross sections as functions of the couplings $g_{H_2VV}$ (left) and $g_{H_2H_1H_1}$ (right). The point on a line marks the actual value of the plotted coupling for the corresponding BP. See text for more details.}
\end{figure}

In Figs.\,\ref{fig:N2-params-1} and \ref{fig:N2-params-2} we show the four cross sections as functions of the most important $H_2$ couplings in this context, for all the six BPs. The former corresponds to the couplings $g_{H_2 t\bar{t}}$ and $g_{H_2 b\bar{b}}$, and the latter to $g_{H_2VV}$ and $g_{H_2H_1H_1}$, while rows 1, 2, 3 and 4 in both the figures depict $\sigma_{2\times 2}$, $\sigma_{\rm b}$, $\sigma_{\rm c}$ and $R_\sigma$, respectively. The plotted ranges of the couplings are indicative of those observed across all the 100 points obtained for the $m_{H_2}=m_{H_3}=410$\,GeV scenario. Once again, for each BP in a given panel, all the remaining couplings are fixed to the values described in Table\,\ref{tab:N2-BPs}. The red lines in the figure correspond to BP1, green to BP2, olive to BP3, violet to BP4, blue to BP5 and orange to BP6. The point on a line in a given panel marks the actual value of the plotted coupling for that BP. The horizontal black lines in the panels in columns 1, 2 and 3, indicate the current experimental limit on \hobs\ pair-production cross section\,\cite{Sirunyan:2018ayu}, which we approximate to be 1\,pb for $m_{H_2} =m_{H_3}=410$\,GeV considered here. Note also that, since only the product of the corresponding $H_2$ and $H_3$ couplings enters the $i{{\mathfrak{I}}{\rm m}\hat\Pi}_{23}(\hat s)$ element of the Higgs propagator matrix, the behaviour of $\sigma_{\rm c}$ with varying $H_3$ couplings should by and large mimic that with varying $H_2$ couplings.

In the left column of the Fig.\,\ref{fig:N2-params-1}, one sees that the presence of an additional Higgs boson degenerate in mass with $H_2$ eliminates the peaks appearing at specific values of $g_{H_2 t\bar{t}}$ in $\sigma_{2\times 2}$, so that the variations in $\sigma_{\rm b}$ in the second row are smoother than in the first row. As expected, $\sigma_{\rm b}$ shows a very similar behaviour for BP1 and BP2, reaching values even higher than the true ones for slightly different positive $g_{H_2 t\bar{t}}$ (recall that $g_{H_3 t\bar{t}}$ is also positive for these two points). Thus $g_{H_2 t\bar{t}}\gtrsim 0.1$ would be ruled out by the LHC \hobs\hobs-production limits. When $g_{H_2 t\bar{t}}$ switches sign to negative, $\sigma_{\rm b}$ drops to much smaller values. The introduction of the full propagator matrix then largely mitigates the very strong dependence of the cross section on positive $g_{H_3 t\bar{t}}$, as seen in the third row, bringing it down to values consistent with experimental bounds. And since $\sigma_{\rm c}$ shows little variation with $g_{H_2 t\bar{t}}$, the shapes of the red and green lines in the bottom row of this column (and also of the right column) are very similar to those in row 2, with $R_\sigma$ reaching about 900 for the BP1.

Cross sections for BP3 and BP4, both of which have $g_{H_3 t\bar{t}}$ with mutually opposite signs but very similar magnitudes, show similar trends to each other with the variations in $g_{H_2 t\bar{t}}$ across the four panels on the left. For these two points, the peaks in $\sigma_{2\times 2}$ are the tallest, while $\sigma_{\rm b}$ and even $\sigma_{\rm c}$ violates the experimental bound for large negative $g_{H_2 t\bar{t}}$. BP4 and BP5, likewise mimic each other's behaviour for positive $g_{H_2 t\bar{t}}$, but since BP4 has a negative $g_{H_3 t\bar{t}}$ larger in magnitude than that in BP5, its dependence on negative $g_{H_2 t\bar{t}}$ is much more pronounced for both $\sigma_{\rm b}$ and $\sigma_{\rm c}$. The right column of the figure shows negligible dependence of $\sigma_b$ on $g_{H_2 b\bar{b}}$ for all the BPs expect 1 and 2 and, conversely, the least variation in $\sigma_c$ for these two BPs. This is due to the fact that for these points $g_{H_2 b\bar{b}}$ and $g_{H_3 b\bar{b}}$ both have negative signs, opposite to the signs of the two top-Yukawa couplings which have a much more dominant effect. 

The left column of Fig.\,\ref{fig:N2-params-2} illustrates that the $g_{H_2VV}$ coupling plays a role as crucial as the top-Yukawa couplings. Similarly to the NMSSM, these couplings originally have generally quite small magnitudes, as a consequence of the very SM-like properties of the $H_1$. For this coupling, the two peaks appearing in $\sigma_{2\times 2}$ are replaced by a tall narrow peak in $\sigma_{\rm b}$, close to $g_{H_2VV}=0$. The introduction of the full propagator matrix brings even the highest of all the peak values of $\sigma_{\rm b}$, seen for BP1 and BP2, down to an experimentally acceptable sub-pb level. The shapes of all the lines are hence largely dictated by the interplay between the signs and sizes of the top-Yukawa and gauge couplings of $H_2$ and $H_3$. In the right column is depicted the dependence of the cross sections on $g_{H_2 H_1H_1}$, which is the only coupling of significance other than the ones discussed above. Here, $\sigma_{2\times 2}$ shows a sharp dip at the zero of this coupling, since it also enters the $H_2\to H_1H_1$ decay besides the self-energies. This sharp dip shifts away from zero for $\sigma_{\rm b}$, according to the relative sign of the diagonal $H_3$ contributions to the propagator. It returns to zero when the off-diagonal terms are also turned on. Around the minimum, $\sigma_{\rm c}$ shows a fairly symmmetric behaviour in both signs of $g_{H_2 H_1H_1}$, as do $\sigma_{\rm b}$ and $\sigma_{\rm b}$. Unlike these two cross sections, however, $\sigma_{\rm c}$ increases rather smoothly. 

Finally, a negligible dependence of each of the cross sections on all of the remaining couplings given in Table\,\ref{tab:N2-BPs} was noted, since the corresponding particle pairs are rather heavy. The contribution to the Higgs self-energies from even the relatively lighter pairs, such as $AZ$ and $H^+ W^-$, for any value of the coupling is vanishing. Plots illustrating variations in the rest of the couplings can therefore be safely dropped. 

\section{Conclusions}
\label{sec:concl}

The commonly adopted approach of calculating the cross section for a given $2\to 2$ process by factorising it into its production and decay parts cannot account for possible quantum interference amongst the propagators of several mass-degenerate states. This is because such an approach by construction assumes narrow widths for the resonant mediators. In some previous papers we explored such interference effects in the case of the gluon-fusion production of certain SM final states, via two highly mass-degenerate Higgs mediators. The mass-splitting between the two intermediate Higgs states being comparable to or smaller than the sum of their widths is a precondition for such effects to be sizeable, in both the integral and the differential cross section. The reason for their onset is that the imaginary off-diagonal elements of the Higgs propagator matrix become comparable to the imaginary parts of the diagonal elements, irrespectively of whether they are taken into account coherently or incoherently. These studies were performed within the illustrative theoretical framework of the NMSSM.

In this article, we have extended our investigation to the pair-production of the lightest of the three NMSSM neutral Higgs scalars, $H_1$, at the 14\,TeV LHC, taking into account the contributions of the triangle as well as the box diagram that this process proceeds through. We have investigated the impact of not only the interference between these two topologies, but also of the aforementioned propagator interference between $H_2$ and $H_3$ within the triangle topology on the cross section for $H_1H_1$ production. Furthermore, since the lowest $H_3$ mass attainable in the NMSSM is $405$\,GeV, owing to the bounds from the LHC searches, we have also included the N2HDM in our analysis. In this framework, since the physical Higgs boson masses are input parameters, we could choose any desired (unique) value for $m_{H_2}$ and $m_{H_3}$, which allowed us to study also the scenario where they contribute non-resonantly to the triangle topology for the studied process. 

In the case of the NMSSM, we have found the effects of the inclusion of the full Higgs propagator matrix in the triangle topology to be similar in size to those established in our previous studies. However, given the various constraints imposed, since the minimal mass-splitting between $H_2$ and $H_3$ is obtained in a very narrow region of the parameter space where, however, the sum of their widths never exceeds it, the effect is largely subdued. In this region the box diagram and the triangle diagram with (an off-shell) $H_1$ in the propagator contribute much more dominantly to the $H_1$ pair-production process. The narrowness of this region also means that a nearly constant negative interference is always observed between the two topologies. 

In the N2HDM, on the other hand, we have seen that the propagator interference effects can modify the cross section by more than two orders of magnitude. Of particular importance is the observation that these effects tend to `regulate' the behavior of the total $H_1H_1$ cross section, smoothing the peaks that appear in it for certain specific values of the $H_2$ and $H_3$ couplings, and generally bringing it down to values consistent with the current LHC limits. Moreover, herein the interference between the box and triangle topologies can be positive or negative, as a consequence of the relatively wider ranges of the magnitudes and sign combinations of the Higgs boson Yukawa couplings. Clearly, such a disparity between the results obtained for this model and those for the NMSSM is due to the fact that supersymmetry imposes strong limitations on the masses and couplings of the heavy Higgs states. In the N2HDM, these quantities are essentially free parameters. But even in this model, when $m_{H_2}$ and $m_{H_3}$ lie below the $H_1H_1$ production threshold, the propagator interference effects tend to vanish. In this case the one-loop two-point functions corresponding to the off-diagonal elements in the Higgs propagator matrix are too small to be able to overcome the kinematic suppression. 

Finally, we emphasise that we have reached the above conclusions on the basis of a detailed analysis at the level of the total cross section. As for their phenomenological relevance, the LHC may develop sensitivity to all such interesting dynamics already at its upcoming Run 3, at least in the N2HDM. Hence, for the purpose of aiding experimental efforts in establishing all the effects studied here, we have proposed some BPs,  compliant with the latest theoretical and experimental constraints, that are amenable to dedicated probes by the ATLAS and CMS collaborations.  


\section*{Acknowledgments}

BD acknowledges the financial support provided by ICTP-EAIFR, where part of this project was carried out. SMo is supported in part through the NExT Institute and STFC Consolidated Grant ST/L000296/1. PP thanks the Department of Physics, Concordia University, for its hospitality during the later part of this project. 


\begin{thebibliography}{10}

\bibitem{Aad:2012tfa}
{\scshape ATLAS Collaboration}, G.~Aad et~al., \emph{{Observation
  of a new particle in the search for the Standard Model Higgs boson with the
  ATLAS detector at the LHC}},
  \href{http://dx.doi.org/10.1016/j.physletb.2012.08.020}{\emph{Phys.Lett.}
  {\bf B716} (2012) 1}, [\href{http://arxiv.org/abs/1207.7214}{{\tt
  1207.7214}}].

\bibitem{Chatrchyan:2012xdj}
{\scshape CMS Collaboration}, S.~Chatrchyan et~al.,
  \emph{{Observation of a new boson at a mass of 125 GeV with the CMS
  experiment at the LHC}},
  \href{http://dx.doi.org/10.1016/j.physletb.2012.08.021}{\emph{Phys. Lett.}
  {\bf B716} (2012) 30}, [\href{http://arxiv.org/abs/1207.7235}{{\tt
  1207.7235}}].

\bibitem{Branco:2011iw}
G.~C. Branco, P.~M. Ferreira, L.~Lavoura, M.~N. Rebelo, M.~Sher and J.~P.
  Silva, \emph{{Theory and phenomenology of two-Higgs-doublet models}},
  \href{http://dx.doi.org/10.1016/j.physrep.2012.02.002}{\emph{Phys. Rept.}
  {\bf 516} (2012) 1}, [\href{http://arxiv.org/abs/1106.0034}{{\tt
  1106.0034}}].

\bibitem{Gunion:1989we}
J.~F. Gunion, H.~E. Haber, G.~L. Kane and S.~Dawson, \emph{{The Higgs Hunter's
  Guide}}, {\emph{Front. Phys.} {\bf 80} (2000) 1}.

\bibitem{Gunion:1992hs}
J.~F. Gunion, H.~E. Haber, G.~L. Kane and S.~Dawson, \emph{{Errata for the
  Higgs hunter's guide}}, \href{http://arxiv.org/abs/hep-ph/9302272}{{\tt
  hep-ph/9302272}}.

\bibitem{Djouadi:2005gj}
A.~Djouadi, \emph{{The Anatomy of electro-weak symmetry breaking. II. The Higgs
  bosons in the minimal supersymmetric model}},
  \href{http://dx.doi.org/10.1016/j.physrep.2007.10.005}{\emph{Phys. Rept.}
  {\bf 459} (2008) 1}, [\href{http://arxiv.org/abs/hep-ph/0503173}{{\tt
  hep-ph/0503173}}].
  
\bibitem{Gunion:2002zf}
J.~F. Gunion and H.~E. Haber, \emph{{The CP conserving two Higgs doublet model:
  The Approach to the decoupling limit}},
  \href{http://dx.doi.org/10.1103/PhysRevD.67.075019}{\emph{Phys. Rev. D} {\bf
  67} (2003) 075019}, [\href{http://arxiv.org/abs/hep-ph/0207010}{{\tt
  hep-ph/0207010}}].

\bibitem{Carena:2013ooa}
M.~Carena, I.~Low, N.~R. Shah and C.~E.~M. Wagner, \emph{{Impersonating the
  Standard Model Higgs Boson: Alignment without Decoupling}},
  \href{http://dx.doi.org/10.1007/JHEP04(2014)015}{\emph{JHEP} {\bf 04} (2014)
  015}, [\href{http://arxiv.org/abs/1310.2248}{{\tt 1310.2248}}].

\bibitem{Haber:2017erd}
H.~E. Haber, S.~Heinemeyer and T.~Stefaniak, \emph{{The Impact of Two-Loop
  Effects on the Scenario of MSSM Higgs Alignment without Decoupling}},
  \href{http://dx.doi.org/10.1140/epjc/s10052-017-5243-5}{\emph{Eur. Phys. J.
  C} {\bf 77} (2017) 742}, [\href{http://arxiv.org/abs/1708.04416}{{\tt
  1708.04416}}].

\bibitem{Ellis:2004fs}
J.~R. Ellis, J.~S. Lee and A.~Pilaftsis, \emph{{CERN LHC signatures of resonant
  CP violation in a minimal supersymmetric Higgs sector}},
  \href{http://dx.doi.org/10.1103/PhysRevD.70.075010}{\emph{Phys.Rev.} {\bf
  D70} (2004) 075010}, [\href{http://arxiv.org/abs/hep-ph/0404167}{{\tt
  hep-ph/0404167}}].

\bibitem{Fuchs:2014ola}
E.~Fuchs, S.~Thewes and G.~Weiglein, \emph{{Interference effects in BSM
  processes with a generalised narrow-width approximation}},
  \href{http://dx.doi.org/10.1140/epjc/s10052-015-3472-z}{\emph{Eur. Phys. J.}
  {\bf C75} (2015) 254}, [\href{http://arxiv.org/abs/1411.4652}{{\tt
  1411.4652}}].

\bibitem{Fuchs:2016swt}
E.~Fuchs and G.~Weiglein, \emph{{Breit-Wigner approximation for propagators of
  mixed unstable states}},
  \href{http://dx.doi.org/10.1007/JHEP09(2017)079}{\emph{JHEP} {\bf 09} (2017)
  079}, [\href{http://arxiv.org/abs/1610.06193}{{\tt 1610.06193}}].

\bibitem{Fuchs:2017wkq}
E.~Fuchs and G.~Weiglein, \emph{{Impact of CP-violating interference effects on
  MSSM Higgs searches}},
  \href{http://dx.doi.org/10.1140/epjc/s10052-018-5543-4}{\emph{Eur. Phys. J.}
  {\bf C78} (2018) 87}, [\href{http://arxiv.org/abs/1705.05757}{{\tt
  1705.05757}}].

\bibitem{Fayet:1974pd}
P.~Fayet, \emph{{Supergauge Invariant Extension of the Higgs Mechanism and a
  Model for the electron and Its Neutrino}},
  \href{http://dx.doi.org/10.1016/0550-3213(75)90636-7}{\emph{Nucl.Phys.} {\bf
  B90} (1975) 104}.

\bibitem{Ellis:1988er}
J.~R. Ellis, J.~Gunion, H.~E. Haber, L.~Roszkowski and F.~Zwirner, \emph{{Higgs
  Bosons in a Nonminimal Supersymmetric Model}},
  \href{http://dx.doi.org/10.1103/PhysRevD.39.844}{\emph{Phys.Rev.} {\bf D39}
  (1989) 844}.

\bibitem{Durand:1988rg}
L.~Durand and J.~L. Lopez, \emph{{Upper Bounds on Higgs and Top Quark Masses in
  the Flipped SU(5) x U(1) Superstring Model}},
  \href{http://dx.doi.org/10.1016/0370-2693(89)90079-8}{\emph{Phys.Lett.} {\bf
  B217} (1989) 463}.

\bibitem{Drees:1988fc}
M.~Drees, \emph{{Supersymmetric Models with Extended Higgs Sector}},
  \href{http://dx.doi.org/10.1142/S0217751X89001448}{\emph{Int.J.Mod.Phys.}
  {\bf A4} (1989) 3635}.

\bibitem{Das:2017tob}
B.~Das, S.~Moretti, S.~Munir and P.~Poulose, \emph{{Two Higgs bosons near 125
  GeV in the NMSSM: beyond the narrow width approximation}},
  \href{http://dx.doi.org/10.1140/epjc/s10052-017-5096-y}{\emph{Eur. Phys. J.}
  {\bf C77} (2017) 544}, [\href{http://arxiv.org/abs/1704.02941}{{\tt
  1704.02941}}].

\bibitem{Das:2018haz}
B.~Das, S.~Moretti, S.~Munir and P.~Poulose, \emph{{Quantum interference among
  heavy NMSSM Higgs bosons}},
  \href{http://dx.doi.org/10.1103/PhysRevD.98.055020}{\emph{Phys. Rev.} {\bf
  D98} (2018) 055020}, [\href{http://arxiv.org/abs/1804.10393}{{\tt
  1804.10393}}].

\bibitem{Plehn:1996wb}
T.~Plehn, M.~Spira and P.~Zerwas, \emph{{Pair production of neutral Higgs
  particles in gluon-gluon collisions}},
  \href{http://dx.doi.org/10.1016/0550-3213(96)00418-X}{\emph{Nucl. Phys. B}
  {\bf 479} (1996) 46}, [\href{http://arxiv.org/abs/hep-ph/9603205}{{\tt
  hep-ph/9603205}}].

\bibitem{Lee:2003nta}
J.~Lee, A.~Pilaftsis, M.~S. Carena, S.~Choi, M.~Drees et~al., \emph{{CPsuperH:
  A Computational tool for Higgs phenomenology in the minimal supersymmetric
  standard model with explicit CP violation}},
  \href{http://dx.doi.org/10.1016/S0010-4655(03)00463-6}{\emph{Comput.Phys.Commun.}
  {\bf 156} (2004) 283}, [\href{http://arxiv.org/abs/hep-ph/0307377}{{\tt
  hep-ph/0307377}}].

\bibitem{Baglio:2013iia}
J.~Baglio, R.~Grober, M.~Muhlleitner, D.~Nhung, H.~Rzehak et~al.,
  \emph{{NMSSMCALC: A Program Package for the Calculation of Loop-Corrected
  Higgs Boson Masses and Decay Widths in the (Complex) NMSSM}},
  \href{http://dx.doi.org/10.1016/j.cpc.2014.08.005}{\emph{Comput.Phys.Commun.}
  {\bf 185} 12 (2014) 3372}, [\href{http://arxiv.org/abs/1312.4788}{{\tt 1312.4788}}].

\bibitem{Carena:2000yi}
M.~S. Carena, J.~R. Ellis, A.~Pilaftsis and C.~Wagner, \emph{{Renormalization
  group improved effective potential for the MSSM Higgs sector with explicit CP
  violation}},
  \href{http://dx.doi.org/10.1016/S0550-3213(00)00358-8}{\emph{Nucl.Phys.} {\bf
  B586} (2000) 92}, [\href{http://arxiv.org/abs/hep-ph/0003180}{{\tt
  hep-ph/0003180}}].

\bibitem{Cacciapaglia:2009ic}
G.~Cacciapaglia, A.~Deandrea and S.~De~Curtis, \emph{{Nearby resonances beyond
  the Breit-Wigner approximation}},
  \href{http://dx.doi.org/10.1016/j.physletb.2009.10.090}{\emph{Phys. Lett.}
  {\bf B682} (2009) 43}, [\href{http://arxiv.org/abs/0906.3417}{{\tt
  0906.3417}}].

\bibitem{NMSSMTools}
See [\href{http://www.th.u-psud.fr/NMHDECAY/nmssmtools.html}{{\tt http://www.th.u-psud.fr/NMHDECAY/nmssmtools.html}}].

\bibitem{Ellwanger:2004xm}
U.~Ellwanger, J.~F. Gunion and C.~Hugonie, \emph{{NMHDECAY: A Fortran code for
  the Higgs masses, couplings and decay widths in the NMSSM}},
  \href{http://dx.doi.org/10.1088/1126-6708/2005/02/066}{\emph{JHEP} {\bf 0502}
  (2005) 066}, [\href{http://arxiv.org/abs/hep-ph/0406215}{{\tt
  hep-ph/0406215}}].

\bibitem{Ellwanger:2005dv}
U.~Ellwanger and C.~Hugonie, \emph{{NMHDECAY 2.0: An Updated program for
  sparticle masses, Higgs masses, couplings and decay widths in the NMSSM}},
  \href{http://dx.doi.org/10.1016/j.cpc.2006.04.004}{\emph{Comput.Phys.Commun.}
  {\bf 175} (2006) 290}, [\href{http://arxiv.org/abs/hep-ph/0508022}{{\tt
  hep-ph/0508022}}].

\bibitem{Bechtle:2008jh}
P.~Bechtle, O.~Brein, S.~Heinemeyer, G.~Weiglein and K.~E. Williams,
  \emph{{HiggsBounds: Confronting Arbitrary Higgs Sectors with Exclusion Bounds
  from LEP and the Tevatron}},
  \href{http://dx.doi.org/10.1016/j.cpc.2009.09.003}{\emph{Comput.Phys.Commun.}
  {\bf 181} (2010) 138}, [\href{http://arxiv.org/abs/0811.4169}{{\tt
  0811.4169}}].

\bibitem{Bechtle:2011sb}
P.~Bechtle, O.~Brein, S.~Heinemeyer, G.~Weiglein and K.~E. Williams,
  \emph{{HiggsBounds 2.0.0: Confronting Neutral and Charged Higgs Sector
  Predictions with Exclusion Bounds from LEP and the Tevatron}},
  \href{http://dx.doi.org/10.1016/j.cpc.2011.07.015}{\emph{Comput.Phys.Commun.}
  {\bf 182} (2011) 2605}, [\href{http://arxiv.org/abs/1102.1898}{{\tt
  1102.1898}}].

\bibitem{Bechtle:2013wla}
P.~Bechtle, O.~Brein, S.~Heinemeyer, O.~St\aa{}l, T.~Stefaniak et~al.,
  \emph{{$\mathsf{HiggsBounds}-4$: Improved Tests of Extended Higgs Sectors
  against Exclusion Bounds from LEP, the Tevatron and the LHC}},
  \href{http://dx.doi.org/10.1140/epjc/s10052-013-2693-2}{\emph{Eur.Phys.J.}
  {\bf C74} (2014) 2693}, [\href{http://arxiv.org/abs/1311.0055}{{\tt
  1311.0055}}].

\bibitem{Bechtle:2015pma}
P.~Bechtle, S.~Heinemeyer, O.~Stal, T.~Stefaniak and G.~Weiglein,
  \emph{{Applying Exclusion Likelihoods from LHC Searches to Extended Higgs
  Sectors}}, \href{http://dx.doi.org/10.1140/epjc/s10052-015-3650-z}{\emph{Eur.
  Phys. J. C} {\bf 75} (2015) 421},
  [\href{http://arxiv.org/abs/1507.06706}{{\tt 1507.06706}}].

\bibitem{Bechtle:2020pkv}
P.~Bechtle, D.~Dercks, S.~Heinemeyer, T.~Klingl, T.~Stefaniak, G.~Weiglein
  et~al., \emph{{HiggsBounds-5: Testing Higgs Sectors in the LHC 13 TeV Era}},
  \href{http://arxiv.org/abs/2006.06007}{{\tt 2006.06007}}.

\bibitem{vonBuddenbrock:2016rmr}
S.~von Buddenbrock, N.~Chakrabarty, A.~S. Cornell, D.~Kar, M.~Kumar, T.~Mandal
  et~al., \emph{{Phenomenological signatures of additional scalar bosons at the
  LHC}}, \href{http://dx.doi.org/10.1140/epjc/s10052-016-4435-8}{\emph{Eur.
  Phys. J. C} {\bf 76} (2016) 580},
  [\href{http://arxiv.org/abs/1606.01674}{{\tt 1606.01674}}].

\bibitem{Muhlleitner:2016mzt}
M.~M\"uhlleitner, M.~O.~P. Sampaio, R.~Santos and J.~Wittbrodt, \emph{{The N2HDM
  under Theoretical and Experimental Scrutiny}},
  \href{http://dx.doi.org/10.1007/JHEP03(2017)094}{\emph{JHEP} {\bf 03} (2017)
  094}, [\href{http://arxiv.org/abs/1612.01309}{{\tt 1612.01309}}].

\bibitem{Coimbra:2013qq}
R.~Coimbra, M.~O. Sampaio and R.~Santos, \emph{{ScannerS: Constraining the
  phase diagram of a complex scalar singlet at the LHC}},
  \href{http://dx.doi.org/10.1140/epjc/s10052-013-2428-4}{\emph{Eur. Phys. J.
  C} {\bf 73} (2013) 2428}, [\href{http://arxiv.org/abs/1301.2599}{{\tt
  1301.2599}}].

\bibitem{Muhlleitner:2020wwk}
M.~M\"uhlleitner, M.~O. Sampaio, R.~Santos and J.~Wittbrodt, \emph{{ScannerS:
  Parameter Scans in Extended Scalar Sectors}},
  \href{http://arxiv.org/abs/2007.02985}{{\tt 2007.02985}}.

\bibitem{Krause:2017mal}
M.~Krause, D.~Lopez-Val, M.~M\"uhlleitner and R.~Santos, \emph{{Gauge-independent
  Renormalization of the N2HDM}},
  \href{http://dx.doi.org/10.1007/JHEP12(2017)077}{\emph{JHEP} {\bf 12} (2017)
  077}, [\href{http://arxiv.org/abs/1708.01578}{{\tt 1708.01578}}].

\bibitem{Ferreira:2019iqb}
P.~Ferreira, M.~M\"uhlleitner, R.~Santos, G.~Weiglein and J.~Wittbrodt,
  \emph{{Vacuum Instabilities in the N2HDM}},
  \href{http://dx.doi.org/10.1007/JHEP09(2019)006}{\emph{JHEP} {\bf 09} (2019)
  006}, [\href{http://arxiv.org/abs/1905.10234}{{\tt 1905.10234}}].

\bibitem{Sirunyan:2018koj}
{\scshape CMS Collaboration}, A.~M. Sirunyan et~al., \emph{{Combined
  measurements of Higgs boson couplings in proton\textendash{}proton collisions
  at $\sqrt{s}=13\,\text {Te}\text {V} $}},
  \href{http://dx.doi.org/10.1140/epjc/s10052-019-6909-y}{\emph{Eur. Phys. J.
  C} {\bf 79} (2019) 421}, [\href{http://arxiv.org/abs/1809.10733}{{\tt
  1809.10733}}].

\bibitem{Engeln:2018mbg}
I.~Engeln, M.~M\"uhlleitner and J.~Wittbrodt, \emph{{N2HDECAY: Higgs Boson
  Decays in the Different Phases of the N2HDM}},
  \href{http://dx.doi.org/10.1016/j.cpc.2018.07.020}{\emph{Comput. Phys.
  Commun.} {\bf 234} (2019) 256},
  [\href{http://arxiv.org/abs/1805.00966}{{\tt 1805.00966}}].

\bibitem{Bechtle:2013xfa}
P.~Bechtle, S.~Heinemeyer, O.~St\aa{}l, T.~Stefaniak and G.~Weiglein,
  \emph{{$HiggsSignals$: Confronting arbitrary Higgs sectors with measurements
  at the Tevatron and the LHC}},
  \href{http://dx.doi.org/10.1140/epjc/s10052-013-2711-4}{\emph{Eur.Phys.J.}
  {\bf C74} (2014) 2711}, [\href{http://arxiv.org/abs/1305.1933}{{\tt
  1305.1933}}].
  
\bibitem{Bechtle:2020uwn}
P.~Bechtle, S.~Heinemeyer, T.~Klingl, T.~Stefaniak, G.~Weiglein and
  J.~Wittbrodt, \emph{{HiggsSignals-2: Probing new physics with precision Higgs
  measurements in the LHC 13 TeV era}},
  \href{http://arxiv.org/abs/2012.09197}{{\tt 2012.09197}}.

\bibitem{Dawson:1998py}
S.~Dawson, S.~Dittmaier and M.~Spira, \emph{{Neutral Higgs boson pair
  production at hadron colliders: QCD corrections}},
  \href{http://dx.doi.org/10.1103/PhysRevD.58.115012}{\emph{Phys. Rev. D} {\bf
  58} (1998) 115012}, [\href{http://arxiv.org/abs/hep-ph/9805244}{{\tt
  hep-ph/9805244}}].

\bibitem{Grober:2015cwa}
R.~Gr\"ober, M.~M\"uhlleitner, M.~Spira and J.~Streicher, \emph{{NLO QCD
  Corrections to Higgs Pair Production including Dimension-6 Operators}},
  \href{http://dx.doi.org/10.1007/JHEP09(2015)092}{\emph{JHEP} {\bf 09} (2015)
  092}, [\href{http://arxiv.org/abs/1504.06577}{{\tt 1504.06577}}].

\bibitem{Agostini:2016vze}
A.~Agostini, G.~Degrassi, R.~Gr\"ober and P.~Slavich, \emph{{NLO-QCD
  corrections to Higgs pair production in the MSSM}},
  \href{http://dx.doi.org/10.1007/JHEP04(2016)106}{\emph{JHEP} {\bf 04} (2016)
  106}, [\href{http://arxiv.org/abs/1601.03671}{{\tt 1601.03671}}].

\bibitem{Degrassi:2016vss}
G.~Degrassi, P.~P. Giardino and R.~Gr\"ober, \emph{{On the two-loop virtual QCD
  corrections to Higgs boson pair production in the Standard Model}},
  \href{http://dx.doi.org/10.1140/epjc/s10052-016-4256-9}{\emph{Eur. Phys. J.
  C} {\bf 76} (2016) 411}, [\href{http://arxiv.org/abs/1603.00385}{{\tt
  1603.00385}}].

\bibitem{Bonciani:2018omm}
R.~Bonciani, G.~Degrassi, P.~P. Giardino and R.~Gr\"ober, \emph{{Analytical
  Method for Next-to-Leading-Order QCD Corrections to Double-Higgs
  Production}},
  \href{http://dx.doi.org/10.1103/PhysRevLett.121.162003}{\emph{Phys. Rev.
  Lett.} {\bf 121} (2018) 162003}, [\href{http://arxiv.org/abs/1806.11564}{{\tt
  1806.11564}}].

\bibitem{Chen:2019lzz}
L.-B. Chen, H.~T. Li, H.-S. Shao and J.~Wang, \emph{{Higgs boson pair
  production via gluon fusion at N$^3$LO in QCD}},
  \href{http://dx.doi.org/10.1016/j.physletb.2020.135292}{\emph{Phys. Lett. B}
  {\bf 803} (2020) 135292}, [\href{http://arxiv.org/abs/1909.06808}{{\tt
  1909.06808}}].

\bibitem{Chen:2019fhs}
L.-B. Chen, H.~T. Li, H.-S. Shao and J.~Wang, \emph{{The gluon-fusion
  production of Higgs boson pair: N$^3$LO QCD corrections and top-quark mass
  effects}}, \href{http://dx.doi.org/10.1007/JHEP03(2020)072}{\emph{JHEP} {\bf
  03} (2020) 072}, [\href{http://arxiv.org/abs/1912.13001}{{\tt 1912.13001}}].

\bibitem{Baglio:2020ini}
J.~Baglio, F.~Campanario, S.~Glaus, M.~M\"uhlleitner, J.~Ronca, M.~Spira
  et~al., \emph{{Higgs-Pair Production via Gluon Fusion at Hadron Colliders:
  NLO QCD Corrections}},
  \href{http://dx.doi.org/10.1007/JHEP04(2020)181}{\emph{JHEP} {\bf 04} (2020)
  181}, [\href{http://arxiv.org/abs/2003.03227}{{\tt 2003.03227}}].

\bibitem{Batell:2015koa}
B.~Batell, M.~McCullough, D.~Stolarski and C.~B. Verhaaren, \emph{{Putting a
  Stop to di-Higgs Modifications}},
  \href{http://dx.doi.org/10.1007/JHEP09(2015)216}{\emph{JHEP} {\bf 09} (2015)
  216}, [\href{http://arxiv.org/abs/1508.01208}{{\tt 1508.01208}}].

\bibitem{Huang:2017nnw}
P.~Huang, A.~Joglekar, M.~Li and C.~E.~M. Wagner, \emph{{Corrections to
  di-Higgs boson production with light stops and modified Higgs couplings}},
  \href{http://dx.doi.org/10.1103/PhysRevD.97.075001}{\emph{Phys. Rev. D} {\bf
  97} (2018) 075001}, [\href{http://arxiv.org/abs/1711.05743}{{\tt
  1711.05743}}].

\bibitem{Huang:2019bcs}
P.~Huang and Y.~H. Ng, \emph{{Di-Higgs Production in SUSY models at the LHC}},
  \href{http://dx.doi.org/10.1140/epjp/s13360-020-00677-1}{\emph{Eur. Phys. J.
  Plus} {\bf 135} (2020) 660}, [\href{http://arxiv.org/abs/1910.13968}{{\tt
  1910.13968}}].

\bibitem{Arbey:2017gmh}
A.~Arbey, F.~Mahmoudi, O.~Stal and T.~Stefaniak, \emph{{Status of the Charged
  Higgs Boson in Two Higgs Doublet Models}},
  \href{http://dx.doi.org/10.1140/epjc/s10052-018-5651-1}{\emph{Eur. Phys. J.
  C} {\bf 78} (2018) 182}, [\href{http://arxiv.org/abs/1706.07414}{{\tt
  1706.07414}}].

\bibitem{Sirunyan:2018ayu}
{\scshape CMS collaboration}, A.~M. Sirunyan et~al., \emph{{Combination of
  searches for Higgs boson pair production in proton-proton collisions at
  $\sqrt{s} = $ 13 TeV}},
  \href{http://dx.doi.org/10.1103/PhysRevLett.122.121803}{\emph{Phys. Rev.
  Lett.} {\bf 122} (2019) 121803}, [\href{http://arxiv.org/abs/1811.09689}{{\tt
  1811.09689}}].

\end{thebibliography}

\providecommand{\href}[2]{#2}\begingroup\raggedright\endgroup

\end{document}